\documentclass[10pt,aps,prb,twocolumn,superscriptaddress,floatfix]{revtex4-2}
\usepackage[colorlinks=true,citecolor=blue,linkcolor=blue,breaklinks=true]{hyperref}

\usepackage{dsfont}

\usepackage{color,graphicx}
\usepackage{bm}
\usepackage{amsmath}
\usepackage{amssymb}
\usepackage{physics}
\usepackage[capitalise]{cleveref}
\allowdisplaybreaks

\begin{document}

\def\afflux{Department of Physics and Materials Science, University of Luxembourg,\\ 1511 Luxembourg, Luxembourg}

\title{Perfectly localized Majorana corner modes in fermionic lattices}
\author{Prathyush P. Poduval}
\email{ppoduval@umd.edu}
\affiliation{Condensed Matter Theory Center, Department of Physics, University of Maryland, College Park, MD 20742, USA}
\affiliation{\afflux}

\author{Thomas L. Schmidt}
\email{thomas.schmidt@uni.lu}
\affiliation{\afflux}
\affiliation{School of Chemical and Physical Sciences, Victoria University of Wellington,\\ P.O.~Box 600, Wellington 6140, New Zealand}

\author{Andreas Haller}
\email{andreas.haller@uni.lu}
\affiliation{\afflux}

\date{\today}

\begin{abstract}
    Focusing on examples of Majorana zero modes on the corners of a two-dimensional lattice, we introduce a method to find parameter regions where the Majorana modes are perfectly localized on a single site.
    Such a limit allows us to study the dimerization structure of the sparse bulk Hamiltonian that results in the higher-order topology of the system.
    Furthermore, such limits typically provide an analytical understanding of the system energy scales.
    Based on the dimerization structure we extract from the two-dimensional model, we identify a more general stacking procedure to construct Majorana zero modes in arbitrary corners of a $d$-dimensional hypercube, which we demonstrate explicitly in $d\leq3$.
\end{abstract}

\maketitle

\section{Introduction}

There have been recent attempts to realize Majorana bound states as zero-dimensional, topological bound states of higher-order (HO) topological insulators (TI)~\cite{phong2017majorana,PhysRevResearch.2.032068,ryu2010topological,laubscher2020kramers,laubscher2020majorana,laubscher2021majorana}.
While normal $d$-dimensional TIs are insulating in the bulk but host $(d-1)$-dimensional surface or edge states~\cite{hasan2010colloquium,chiu2016classification,moore2010birth}, $d$-dimensional HOTIs host $(d-D)$-dimensional states with $D > 1$~\cite{benalcazar2017electric,benalcazar2017quantized,imhof2018topolectrical,geier2018second,schindler2018higher,benalcazar2014classification,song2017d}.
A 2D second-order topological insulator (SOTI), for instance, has insulating bulk and edges but zero-energy corner states.
An established way to generate HOTIs relies on crystalline symmetries, in which case the boundary states are protected by a combination of intrinsic and crystalline symmetries~\cite{ryu2010topological,schnyder2008classification,fu2011topological,zhang2019second}.
One typically obtains a HOTI by breaking certain crystalline symmetries of a TI so that its boundary states localize onto a lower-dimensional manifold~\cite{nguyen2022corner,zhang2021intrinsic,PhysRevResearch.2.032068,pan2021braiding}.
For example, on a cube, the surface states can localize onto the hinges or corners~\cite{schindler2018higher} and edge states of a 2D square lattice can localize onto the corners~\cite{PhysRevResearch.2.032068}.

We consider particle-hole symmetric models on a square lattice hosting a pair of helical edge states.
By introducing an in-plane magnetic field these edge states become gapped, and pairs of Majorana corner states are formed~\cite{phong2017majorana,PhysRevResearch.2.032068}.
The orientation of the magnetic field is locked to the configuration of the emergent corner modes, so the corner states can be moved by rotating the magnetic field.
We seek an answer to the following question: are models featuring zero-energy Majorana corner modes adiabatically connected to a limit where the zero modes are decoupled from the bulk, and how can we characterize the corresponding topological phase?
In other words, our goal is to find an analog of the ``sweet spot'' of the Kitaev chain, where the Majorana bound states are located on different unit cells and the couplings between Majorana modes vanish even in the limit of a finite system. The corresponding dimerization structure readily reveals the geometric interpretation of the pairing between Majorana sites, the roles played by different couplings in the fermionic language, and possible ways to characterize the topological phases~\cite{benalcazar2017electric,benalcazar2017quantized,ezawa2018higher,laubscher2019fractional,laubscher2023fractional}.
Such an understanding makes it possible to identify and propose additional couplings which enhance the robustness of topological Majorana states.
Lastly, the topological sweet spot we unveil in a particular example model leads to a geometric pairing strategy that inspires the construction of a family of zero-energy Majorana corner modes embedded in higher dimensional lattices.

We begin by analyzing the model proposed in Ref.~\cite{PhysRevResearch.2.032068}. It hosts Majorana corner states and constitutes a second-order topological insulator, but within its parameter space the corner modes never become perfectly localized. We develop a scheme to find perfectly localized Majorana corner states by considering an enlarged parameter space and show that these corner states are adiabatically connected to those of the original model. Moreover, we show that the perfectly localized bound states result from a dimerization structure of the model expressed in the Majorana basis, which turns out to consist of pairs of coupled Kitaev chains.

We proceed to construct a topological invariant for this extended 2D model and demonstrate the corner-edge correspondence based on adiabatic changes which rotate the corner modes. The topological invariant we propose is based on a nested Pfaffian invariant, which we compute from band structure of a ribbon geometry~\cite{kitaev2001unpaired,budich2013equivalent,peng2017boundary}.

The method we propose for finding perfectly localized Majorana bound states is fairly general. To demonstrate its scope, we go on to propose a three-dimensional model which features two Majorana corner modes on corners of a simple cubic lattice and which results from a suitable stacking of 2D lattices. We demonstrate that this stacking procedure can be applied to construct models with two Majorana modes on arbitrary corners of a hypercube, protected by embedded Kitaev chains in the topological phase.

\section{Construction of localized Majorana bound states} 

We consider a translation invariant, superconducting tight-binding Hamiltonian, which we generically write as $\hat{H}=\sum_{\bm r'\bm r}\bm c_{\bm r'}^\dagger T_{\bm r'-\bm r}\bm c_{\bm r}$.
Here, $\bm c_{\bm r}$ is a Nambu spinor, which consists of electron creation and annihilation operators and includes sublattice degrees of freedom.
Moreover, $T_{\bm r'-\bm r}$ are the matrices that determine the hopping and pairing terms between the sites at $\bm r$ and $\bm r'$.
For illustration purposes, we restrict ourselves to nearest-neighbor hopping on a $d$-dimensional hypercube.
The Hamiltonian then reduces to
\begin{align}
    \label{eq:H}
    \hat{H} &=
    \sum_{\bm r}\left(\sum_{i}\left(\bm c^\dagger_{\bm r + \hat{\bm a}_i}T_{\hat{\bm a}_i} \bm c_{\bm r}  + {\rm h.c.}\right) + \bm c^\dagger_{\bm r}T_{\bm0}\bm c_{\bm r}\right),
\end{align}
where $\{\hat{\bm a}_i\}$ are the primitive lattice vectors between nearest neighbors.
For example, on a 2D square lattice $\hat{\bm a}_i\in\{a \hat{\bm e}_x,a \hat{\bm e}_y\}$ where $a$ is the lattice constant, which we now set to unity.
For simplicity, we assume that $T_{\hat{\bm a}_i}$ are linear functions of the tight-binding parameters of the model.

Denoting by $\ket{0}$ the ground state of $\hat{H}$, a perfectly localized single-particle state in the corner of the hypercube corresponds to $\ket{\psi_{\bm r_c}} = \sum_{\bm r} {\bm \psi}(\bm r, \bm r_c)\cdot \bm c^\dagger_{\bm r} \ket{0}$, with a wavefunction of the form $\bm \psi(\bm r, \bm r_c) = \bm v\delta_{\bm r, \bm r_c}$, where $\bm v=(v_1,\ldots,v_n)^T$ is a vector in the $n$ components of the sublattice degrees of freedom at a unit cell, $\delta$ denotes the Kronecker delta, and $\bm r_c$ is the spatial location of the associated corner.
By requiring this state to be a zero-energy eigenstate of the Hamiltonian in \cref{eq:H}, we obtain a set of equations $T_{\bm 0}\bm v = 0$ and $T_{\hat{\bm a}_i} \bm v=0$ for all $\hat{\bm a}_i$. The existence of a nontrivial null space is not guaranteed in general. However, in the special cases that we will discuss, we find that we require at least a Hamiltonian with direction-anisotropic tight-binding terms.

    Other types of localization can be revealed as well through this procedure.
    For example in a 3D system, if nontrivial solutions of $T_{\hat{\bm e}_x}\bm v=T_{\hat{\bm e}_y}\bm v=T_{\hat{\bm e}_z}\bm v=0$ exist, then the wavefunction $\bm v$ is associated with a corner mode.
    However, we can also identify hinge modes by lifting one of the null space constraints, e.g., isolated modes along the $\hat{\bm e}_z$ hinge may be found by the solutions of $T_{\hat{\bm e}_x}\bm v=T_{\hat{\bm e}_y}\bm v=0$, without imposing $T_{\hat{\bm e}_z}\bm v=0$.
    Similarly, the kernel of a single condition $T_{\hat{\bm e}_x}\bm v=0$ can reveal localized zero energy surface modes on the $yz$ plane.

Finding the nontrivial parameter constellations such that $T_{\bm 0}$ and $T_{\hat{\bm a}_i}$ have $\bm v$ as a common null eigenvector is a computationally difficult task.
To simplify the set of equations, we later use the fact that a system with topological Majorana zero-energy modes has particle-hole symmetry, with the Majorana modes defined as the corresponding particle-hole symmetric states.
It is therefore practical to choose $\bm v$ such that the corresponding wavefunction ${\bm \psi}({\bm r},{\bm r_c})$ transforms trivially under particle-hole symmetry. The operator form of particle-hole symmetry and the associated transformation from Dirac to Majorana fermions depends on the system at hand, which is why we explain this step for a practical example in the next paragraph.

\section{Majorana bound states in adjacent corners}

We apply the method outlined above to find perfectly localized Majorana corner modes on two corners of a square lattice. For this purpose, we propose the following Bloch Hamiltonian,
\begin{align}
{H}_{\bm k} \label{eq:2D_adjacent_corners}
&=
    (t_0+t_x\cos k_x+t_y\cos k_y)\Sigma_{03}\\
&+
    d_{y}\sin k_y \Sigma_{02} +d_{x}\sin k_x\Sigma_{31}\notag\\
&+
    (s_x\cos k_x +s_y\cos k_y)\Sigma_{11}+ b_x\Sigma_{10}+b_y\Sigma_{23},\notag
\end{align}
where $\Sigma_{ij}=\sigma_i\tau_j$, $\sigma_{i}$ and $\tau_{j}$ denote two sets of Pauli matrices in spin and particle-hole space, respectively.
The total Hamiltonian can be written as $\hat H = \sum_{\bm k}\bm c^\dag_{\bm k}H_{\bm k}\bm c^{\vphantom\dag}_{\bm k}$ using the spinors $\bm c_{\bm k}= (c_{\uparrow,\bm k},c_{\downarrow,\bm k},c_{\uparrow,-\bm k}^\dagger, -c_{\downarrow,-\bm k}^\dagger)^T$.
After a transformation from momentum to real space, we identify the hopping matrices
\begin{subequations}
\begin{align}
    T_{\bm 0} &= t_0\Sigma_{03}+b_x\Sigma_{10}+b_y\Sigma_{23},\\
    T_{\hat{\bm e}_x} &= \frac12\left(s_x\Sigma_{11}-id_{x}\Sigma_{31}+t_x\Sigma_{03}\right),\\
    T_{\hat{\bm e}_y} &= \frac12\left(s_y\Sigma_{11}-id_{y}\Sigma_{02}-t_y\Sigma_{03}\right).
\end{align}
\end{subequations}
The model (\ref{eq:2D_adjacent_corners}) is an anisotropic generalization of the Hamiltonian discussed in Ref.~\cite{PhysRevResearch.2.032068}, and has a higher-order topological phase hosting a pair of Majorana modes on adjacent corners of the square lattice.

The particle-hole symmetry of this model is represented by $U_{\mathcal P}=\Sigma_{31}$, i.e., ${H}_{\bm k}=-U_{\mathcal P}{H}_{-\bm k}^*U_{\mathcal P}^\dagger$, and the symmetry operator can be diagonalized as $M_{\mathcal P} U_{\mathcal P} M_{\mathcal P}^T = \Sigma_{00}$ with $M_\mathcal P=\exp[i\frac{\pi}{4}\left(\Sigma_{30}-\Sigma_{03}\right)]\exp[i\frac{\pi}{4}\left(\Sigma_{02}-\Sigma_{10}\right)]$. The corresponding particle-hole symmetric basis in real space defines the Majorana fermions, given by ${\bm m}_{\bm r} = M_{\mathcal P}{\bm c}_{\bm r}$, and satisfies $\{m_{\bm r,i},m_{\bm r,j}\}=\delta_{ij}/2$. In the fermionic basis, the components of these Majorana elements are given by ${\left(\hat{\bm v}_i\right)_j = M_{\mathcal P,ij}}$.

\begin{figure}[t]
    \centering
    \includegraphics[width=\columnwidth]{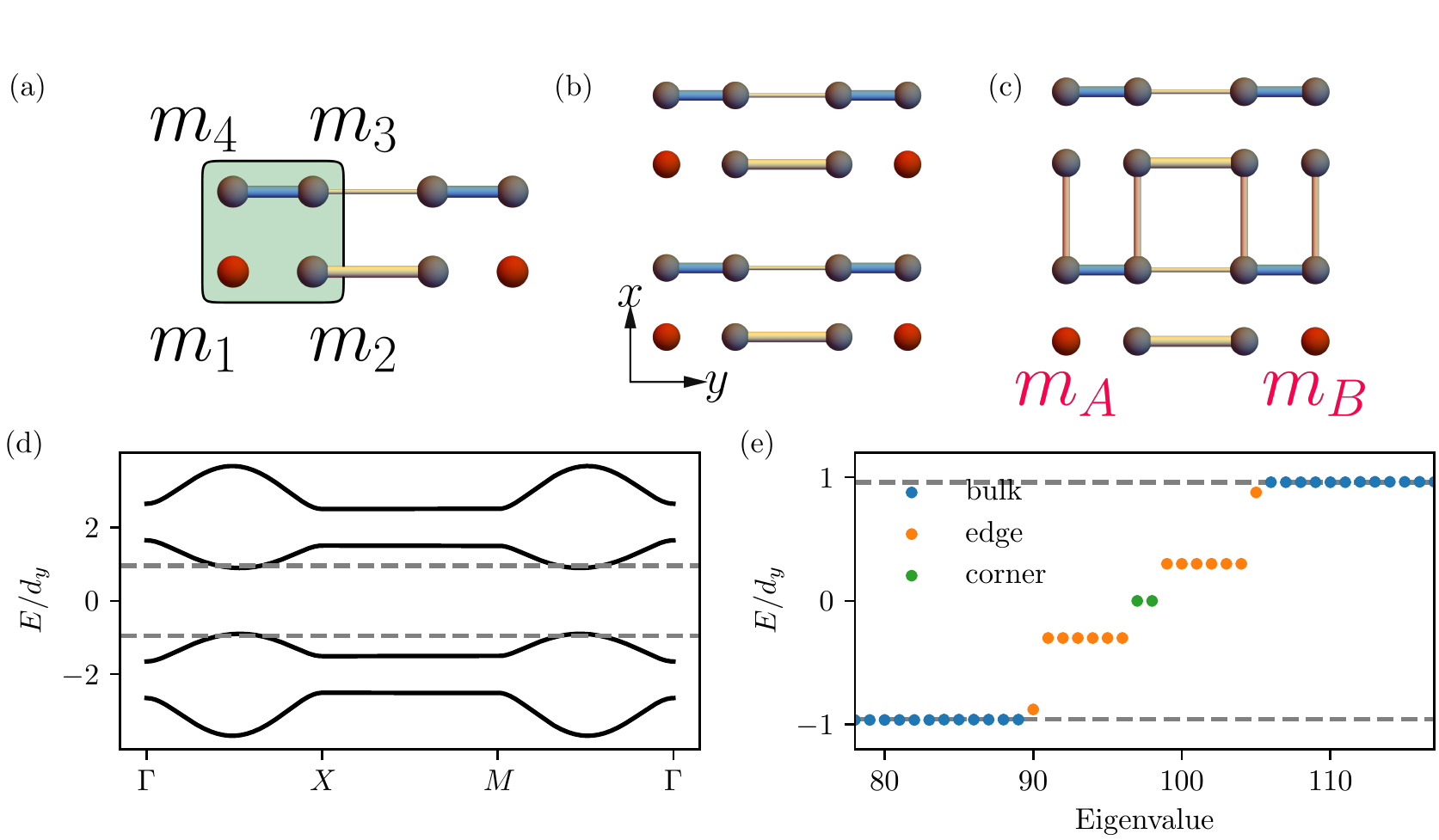}
    \caption{Construction of the dimerization structure in the parameter regimes of \cref{eq:2D_adjacent_corners_parameters}, starting from (a) a pair of topological and trivial Kitaev chains, which are (b) stacked on top of each other and then coupled as in (c) to gap out the intermediary Majorana modes to obtain a 2D bulk model with corner modes in adjacent corners. Panel (d) shows the gapped bulk bands for $t_0=0.5d_y, t_y=0.3d_y, s_x=2d_y$, and panel (e) shows the eigenvalues for a square lattice with $7\times 7$ unit cells with open boundary conditions for the same parameters as in (d). The corner modes (in green) are at zero energy (by construction), and the spectrum shows the gapped bulk (blue) and edge (yellow) bands.}
    \label{fig:2D_adjacent_corner_modes}
\end{figure}

We now apply this technique for revealing Majorana corner modes in the model described by \cref{eq:2D_adjacent_corners}. We consider a square lattice of size $L\times L$ and impose the presence of two Majorana bound modes located on adjacent corner positions $\bm r_1=(0,0)^T$ and $\bm r_2=(0,L)^T$. The corresponding Majorana states are chosen from $\bm v_A,\ \bm v_B\in\{\hat{\bm v}_{1},\hat{\bm v}_{2},\hat{\bm v}_{3},\hat{\bm v}_{4}\}$ defined in the previous paragraph.
To localize the Majorana mode $\bm v_A$ at $\bm r_1$, we impose the conditions $T_{\hat{\bm e}_x}\bm v_A=0$ and $T_{\hat{\bm e}_y}\bm v_A=0$. Similarly, to localize $\bm v_B$ at $\bm r_2$, we require that $T_{\hat{\bm e}_{x}}\bm v_B=0$ and $T_{-\hat{\bm e}_y}\bm v_B=0$.
Solving the four sets of matrix equations simultaneously (two each for $\bm v_A$ and $\bm v_B$) for all possible choices of $\bm v_A$ and $\bm v_B$ gives $16$ distinct solutions, since $\bm v_A$ and $\bm v_B$ can be chosen to belong to one of the four Majorana modes individually.
Some of the solutions correspond to ``trivial'' parameters in the tight-binding model where more than the required corner sites are isolated from all other sites of the lattice. They include the flat-band cases where the bulk or edge sites also decouple from their neighbors, and are at zero energy.
Throughout this work, we are interested in the set of solutions that carry only two zero-energy modes associated with the two chosen corner states $\bm v_A$ and $\bm v_B$.
We find four nontrivial solutions, given by the following constrains between the parameters of \cref{eq:2D_adjacent_corners},
\begin{align}
    t_x=b_x=0,\phantom{=}b_y=p_1 t_0,\phantom{=}d_{y}= p_1 p_2t_y, \hphantom{=}d_{x}=p_1 s_{x},
    \label{eq:2D_adjacent_corners_parameters}
\end{align}
with $p_1,p_2 = \pm 1$. Note that the isolated corner mode solutions do not exist in the isotropic model \cite{PhysRevResearch.2.032068}. We demonstrate that the anisotropic topological limit and the topological phase of the isotropic model are adiabatically connected (see App.~\ref{sec:appB}).
Choosing a different sign for $p_2$ corresponds to transformations of the kind $k_y\to -k_y$, or alternatively exchanging the sublattice flavor of the two Majorana bound states that are localized in the two corners.

In \cref{fig:2D_adjacent_corner_modes}(d), we display the spectrum of the perfectly localized limit, featuring a bulk and edge gap, together with two midgap states exactly at zero energy -- the perfectly localized corner states by construction.
The Majorana pairing structure is presented in \cref{fig:2D_adjacent_corner_modes}(c), where dangling sites, hosting the Majorana zero modes, are highlighted in red.
Along the $x$ direction, \cref{fig:2D_adjacent_corner_modes}(c) readily reveals a decoupled topological Kitaev chain which protects the zero modes even in the presence of couplings along the $y$ direction as long as the edge and bulk gaps remain finite.
The 2D lattice can therefore be understood in terms of repeating coupled trivial and topological Kitaev chains, which are dressed by additional couplings along the $y$ direction. This interpretation is one of the main results of this paper, and leads to the second main result, that the stacking procedure of coupled Kitaev wires can be used to engineer pairs of corner modes with flexible geometrical configurations in $d$-dimensional hypercubic lattices.
To give more specific examples, we now proceed to propose two similar models, the first one hosting Majorana zero modes on opposite corners, and the second one featuring Majorana corner modes embedded in a 3D cube.

\section{Majoranas in opposite corners} 

\begin{figure}[t]
    \centering
    \includegraphics[width=\columnwidth]{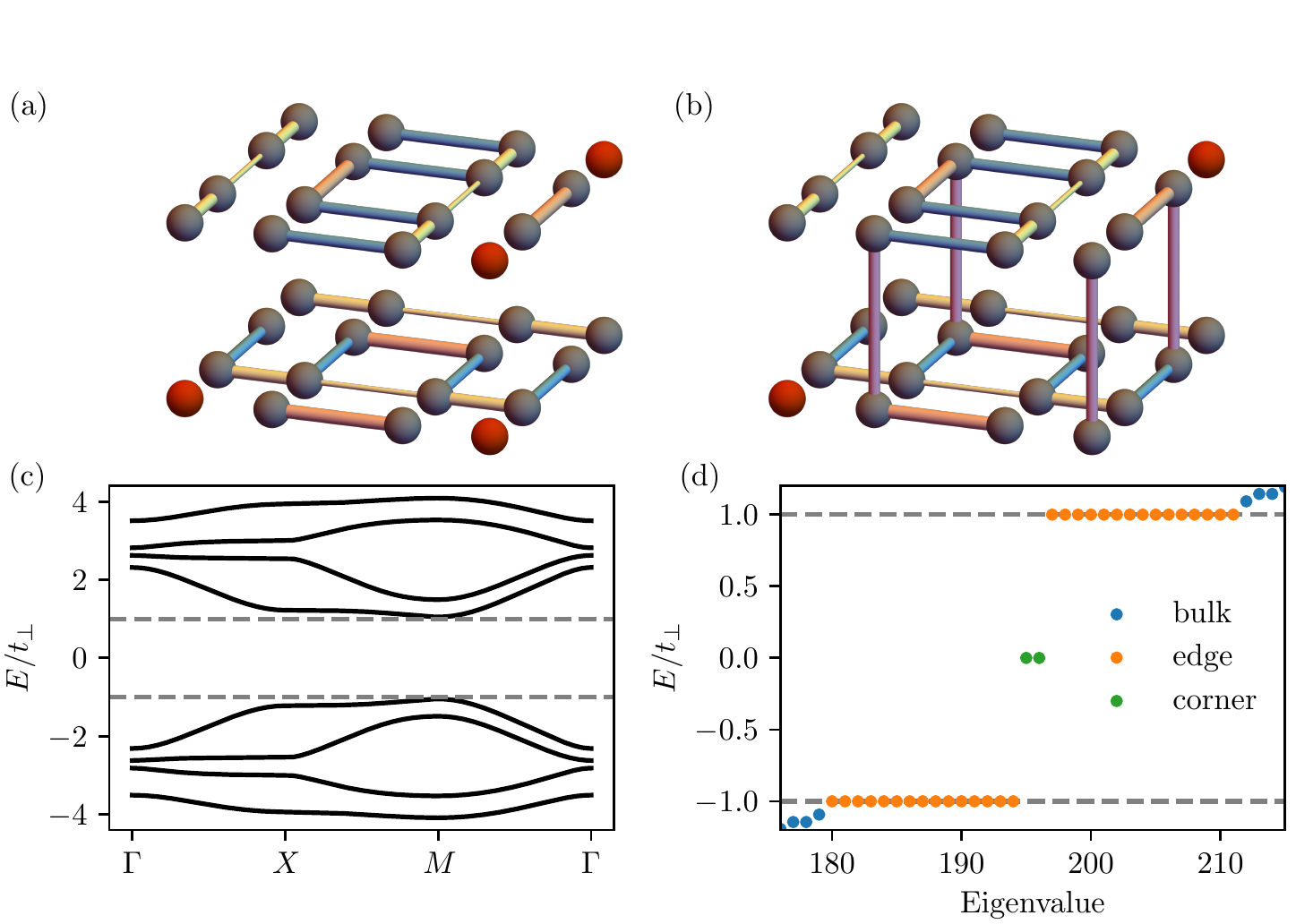}
    \caption{Each of the two sheets in panel (a) features two isolated corner modes. By introducing an additional interlayer coupling [see panel (b)], a pair of corner modes becomes gapped without closing a bulk or edge gap. This results in a 2D bilayer model with isolated Majorana corner modes on opposite corners. In panels (c) and (d), we show the bulk bands and spectrum of a $7\times 7$ lattice for $\mu_1=\mu_2=t_\perp, t_{y,1}=t_{x,2}=2.5t_{\perp},t_{x,1}=t_{y,2}=1.5 t_\perp,t'_{x,1}=t'_{y,2}=\lambda=0$.}
    \label{fig:2D_opposite_corner_modes}
\end{figure}

We proceed to adapt the procedure to construct Majorana modes on opposite corners of a square lattice. Our construction uses two copies of the Hamiltonian~(\ref{eq:2D_adjacent_corners}), with one copy rotated by $90^\circ$ relative to the other one and stacked on top of the other as shown in \cref{fig:2D_opposite_corner_modes}(a).
As a result, a pair of Majorana corner modes overlap on a single corner and form a gapped Dirac mode when additional couplings are introduced (see \cref{fig:2D_opposite_corner_modes}(b)).
The final system, therefore, hosts two diagonally opposite Majorana corner modes.

The associated tight-binding model is given by
\begin{align}\label{eq:bilayer_T}
    T_{\bm 0,1} &= - \mu_{1}\Sigma_{25}, \hphantom{aa}\\
    T_{\bm 0,2} &= -\mu_{2}\Sigma_{25}+\lambda \Sigma_{24},\\
    T_{\hat{\bm e}_y,1} &= \frac{1}{2}t_{y,1}\left(i\Sigma_{02}-\Sigma_{03}\right),\\
    T_{\hat{\bm e}_y,2} &=-i t_{y,2}\Sigma_{-4}-it_{y,2}'\Sigma_{+5},\\
    T_{\hat{\bm e}_x,1}&=-it_{x,1}\Sigma_{+4}-it_{x,1}'\Sigma_{+5},\\
    T_{\hat{\bm e}_x,2} &= \frac12 t_{x,2}\left(i\Sigma_{12}+\Sigma_{13}\right),\\
    T_{\bm 0,12} &= -i t_{\perp}\Sigma_{74},
\end{align}
where again $\Sigma_{ij}=\sigma_i\tau_j$. In addition to the identity matrix $\sigma_0$ and the Pauli matrices $\sigma_{1,2,3}$, we define $\sigma_{\pm} = \frac12\left(\sigma_x\pm i\sigma_y\right)$, $\sigma_{4}=\frac12(\sigma_0+\sigma_1),\sigma_{5}=\frac12(\sigma_0-\sigma_1),\sigma_{6}=\frac12(\sigma_0+\sigma_3),\sigma_{7}=\frac12(\sigma_0-\sigma_3)$. The matrices $\tau_i$ are defined analogously.
Here, $T_{\hat{\bm a}_i,1}$ are the hopping matrices for the first layer, $T_{\hat{\bm a}_i,2}$ are the hopping matrices for the second layer and $T_{\bm 0,12}$ is the hopping between the two layers. The perfectly localized limit in this model corresponds to $\lambda=0$.

The corresponding $8\times 8$ Bloch Hamiltonian ${H}_{\bm k}$ has a particle-hole symmetry represented by $U_{\mathcal P}=\Sigma_{010}$, where $\Sigma_{ijk}=\sigma_i\tau_j\eta_k$ and the Pauli matrices $\eta_k$ act on the layer degree of freedom. $U_{\mathcal P}$ is diagonalized by $M_{\mathcal P} =\left(\Sigma_{000}+i\Sigma_{020}\right)/\sqrt{2}$. \Cref{fig:2D_opposite_corner_modes}(c) shows the spectrum of a finite system with open boundary conditions, where we identify a gapped bulk and two zero-energy states corresponding to the two isolated Majorana modes on adjacent corners.

\section{Nested Pfaffian}
Previous work constructed invariants from the polarization of the model (which are quantized due to crystalline symmetries) to characterize the HOTI state \cite{zhang2020kitaev,ezawa2018higher,ezawa2018minimal,khalaf2018higher,bradlyn2017topological,simon2022higher,trifunovic2021higher,hu2023topological,lenggenhager2022universal}.
In contrast, the only symmetry present in the models considered here is particle-hole symmetry, due to which the periodic BdG Hamiltonian has an associated well-defined Pfaffian~\cite{kitaev2001unpaired,budich2013equivalent}.
Since the corner modes identified in the previous paragraphs are hosted by embedded topological Kitaev chains, a topological classification can be performed through the Pfaffian of the effective edge Hamiltonian.

As an example, we consider a particle-hole symmetric 2D Hamiltonian $H(\bm k)$ that hosts a pair of corner Majorana bound modes. To access the edge Hamiltonian we construct a ribbon geometry, which is infinite in one direction and finite in the other, by a partial Fourier transform to real space,
\begin{align}
    [{H}_y(k_x)]_{y_2,y_1}&=\int \frac{dk_y}{2\pi}{H}_{\bm k} e^{ik_y(y_2-y_1)},\\
    [{H}_x(k_y)]_{x_2,x_1}&=\int \frac{dk_x}{2\pi}{H}_{\bm k} e^{ik_x(x_2-x_1)}.
    \label{eq:ribbondef}
\end{align}
Each effective 1D Hamiltonian describes a pair of edges, which is sufficient for the topological classification of the systems presented here. If the classification of a single edge should be required, we expect that an investigation of semi-infinite domains can be done by constructing the edge Hamiltonian in a similar manner to Ref.~\cite{fidkowski2011model}.
If we consider the spatial indices as internal degrees of freedom, $H_y(k_x)$ and $H_x(k_y)$ represent one-dimensional Bloch Hamiltonians, and the particle hole symmetry passes over from the bulk Hamiltonian as
\begin{align}
    {H}_y(k_x) &= -(U_\mathcal P\otimes \mathds{1}_{y}) {H}^*_y(-k_x) (U_\mathcal{P}^\dagger \otimes\mathds{1}_{y}),\\
    {H}_x(k_y) &= -(U_\mathcal P \otimes  \mathds{1}_{x}) {H}^*_x(-k_y) (U_\mathcal{P}^\dagger\otimes  \mathds{1}_{x}),
\end{align}
where $\mathds{1}_{x}$ and $\mathds{1}_{y}$ are identity operators acting on the real-space indices of the Hamiltonian. In the eigenbasis of the particle-hole symmetry, the one-dimensional Hamiltonians are skew-symmetric and have well-defined Pfaffians, $Q_x = \mathrm{Pf}[{H}_y(k_x)]$ and $Q_y = \mathrm{Pf}[{H}_x(k_y)]$.
Together, the pair ${\bm Q} = (Q_x,Q_y)$ can be used to classify the topology of the edges of a square system, with $\bm Q = (-1,1)$ and $\bm Q = (1,-1)$ meaning that the corner states exist along the two ends of the $x$ ($y$) edge. Moreover, $\bm Q = (-1,-1)$ indicates two non-trivial edges which then support corner modes on opposite corners. $\bm Q = (+1,+1)$ indicates trivial Kitaev chains along both edges, and corresponds to the case of no corner modes.

\section{Pumping corner modes}
In the preceding paragraphs, we constructed two models for inequivalent geometric configurations of two Majorana corner modes on a square lattice -- one displaying them on adjacent corners and the other one on diagonally opposite corners. We can connect them by constructing a pumping protocol in which one of the Majoranas lying in opposite corner is first pumped through the edge and then onto the adjacent layer.
At the end of the pumping cycle, the two Majorana modes will lie on adjacent corners, and the upper layer will be topologically trivial and isolated from the bottom layer. During the process, the edge gap of the upper layer closes to enable transferring the Majorana mode. However, the bulk gap remains nonzero (see App.~\ref{sec:appD}).

\begin{figure}[t]
    \centering
    \includegraphics[width=\columnwidth]{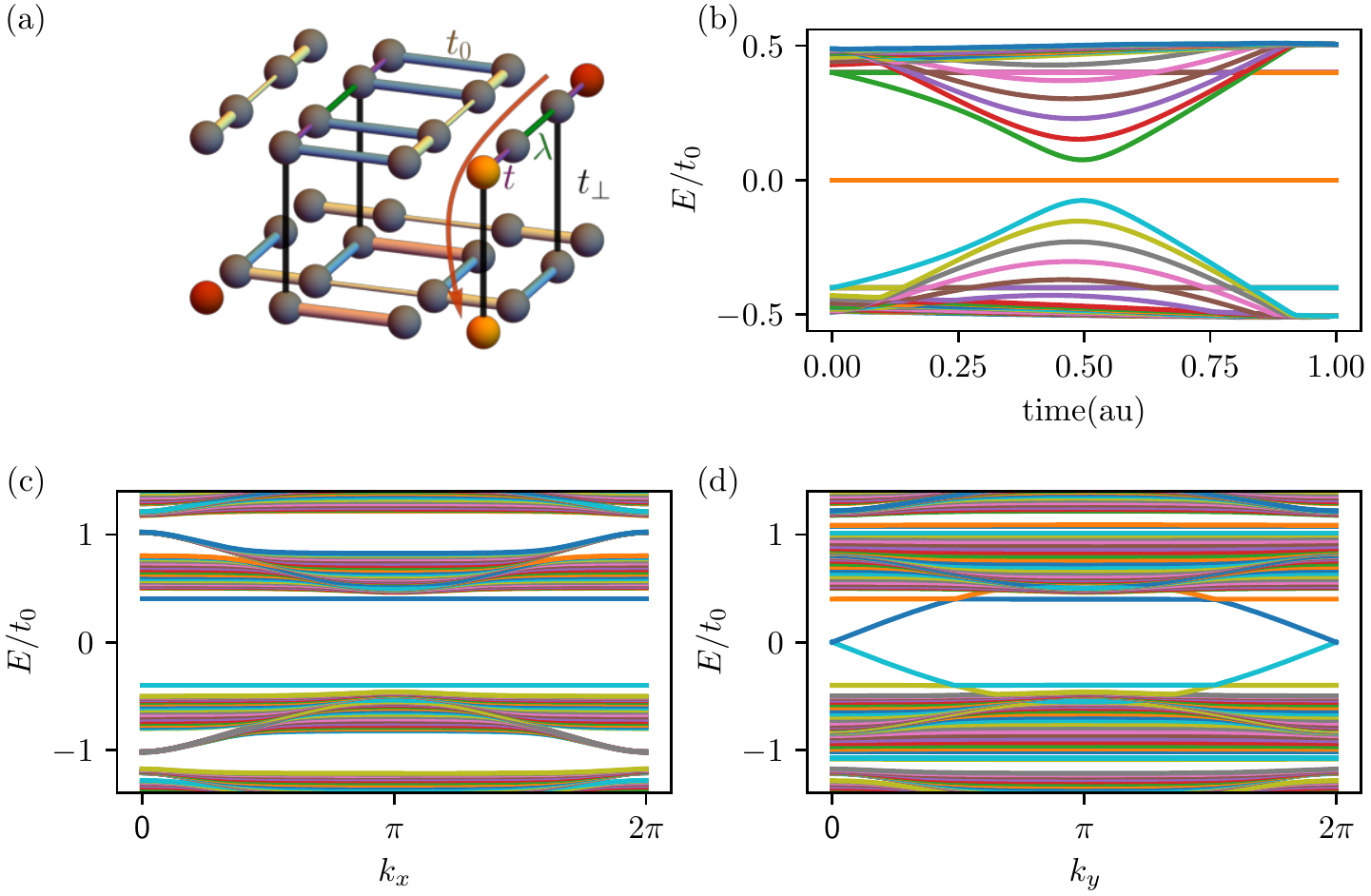}
    \caption{Adiabatic deformation to connect different geometric configurations: moving isolated corner modes from adjacent to opposite corners. (a) The couplings that will be varied (with $t_0$ acting as reference energy scale). (b) The spectrum of the finite system as a function of time, which suggests a gap closing at time $T=0.5$. (c,d) The spectrum of ribbon geometries along the $x$ and $y$ directions, respectively, at time $T=0.5$. This shows that the gap of an edge along the $y$ axis closes at $T=0.5$.}
    \label{fig:pump}
\end{figure}

To pump the Majorana from one corner of the top layer to the adjacent corner of the bottom layer, we need to make the topological Kitaev edge of the top layer trivial, while simultaneously decoupling the two layers.
This is achieved by changing the couplings as shown in \cref{fig:pump}(a),
\begin{align}\label{eq:pump2D}
    t_\perp=0.4(1-T)t_0,\ \lambda=0.6(1-T)t_0,\ t=0.6Tt_0.
\end{align}
\Cref{fig:pump}(b) shows the spectrum of a $10\times 10$ lattice during a pump cycle.
At time $T=0.5$, when the parameters $\lambda=t$, the gap of the finite system appears to close.
This can be better understood by considering the spectra of ribbon geometries along the $x$ and $y$ directions, which we present in \cref{fig:pump}(c) and \cref{fig:pump}(d).
This shows indeed that during the pumping protocol the bulk gap stays finite, but the edge gap closes and reopens, which is expected because the edge Kitaev chain turns from topological to trivial on the top layer, and the associated invariant changes. The bulk spectrum remains gapped throughout the pumping process (see App.~\ref{sec:appD}).

\section{Majorana bound states on the corners of a cube}
\begin{figure}[t]
    \centering
    \includegraphics[width=\columnwidth]{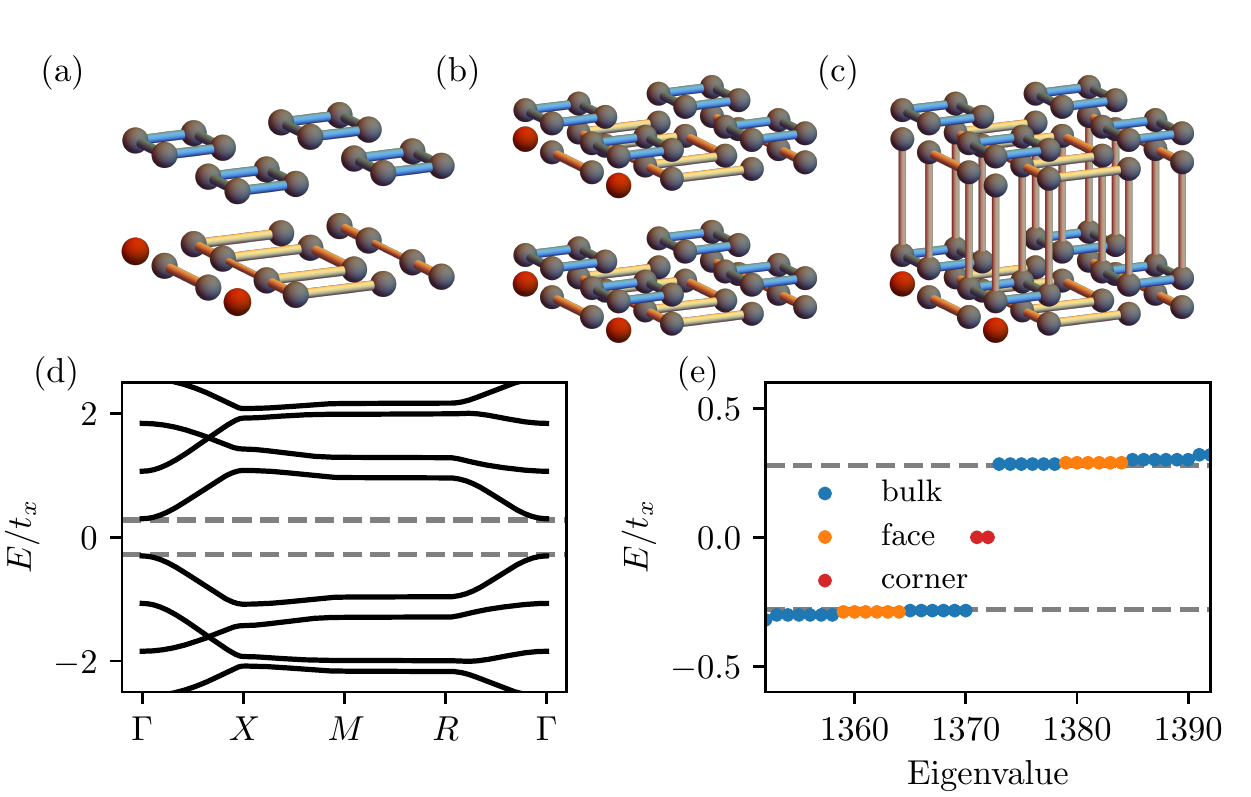}
    \caption{From (a) to (c), we present the strategy to couple a pair of 2D Majorana corner mode models to obtain a 3D bulk model with corner modes in adjacent corners. Panel (d) shows the gapped bulk bands and in panel (e), we show the eigenvalues for a square with $7\times 7\times 7$ unit cells with open boundary conditions. The corner modes (in red) are at zero energy, and the spectrum shows the eigenvalues of clearly gapped bulk (blue) bands. The parameters used are $\alpha_1=-0.9t_x, \mu_1=\mu_2=-0.25t_x,t_{y}=t'_{y}=0.45t_x,t_z=-1.2t_x$.}
    \label{fig:cubeAdjMaj}
\end{figure}

We now demonstrate how to construct models with two localized Majorana modes on adjacent corners in a three-dimensional cubic lattice. To construct the unit cell, we couple a layer with perfectly localized Majorana modes on adjacent corners to a topologically trivial one \cite{matsugatani2018connecting} as shown in \cref{fig:cubeAdjMaj}(a).
Next, the double layer unit cells are repeated along the $z$ axis, see \cref{fig:cubeAdjMaj}(b), and we allow additional tunneling terms to couple the undesired dangling sites to the bulk, see \cref{fig:cubeAdjMaj}(c).
The hopping matrices for this model are given by
\begin{align}
    T_{\bm 0}&=\mu_1 \Sigma_{073}+\mu_2\Sigma_{373}-\alpha_1\Sigma_{720},\\
    T_{\hat{\bm e}_x}&=\frac{t_x}{2}\left(i\Sigma_{662}-\Sigma_{663}\right),\\
    T_{\hat{\bm e}_y}&=\frac{t_{y}}{2}\left(\Sigma_{020}-i\Sigma_{010}\right)+\frac{t_{y}'}{2}\left(\Sigma_{320}-i\Sigma_{310}\right),\\
    T_{\hat{\bm e}_z}&=\frac{t_z}{2}\left(\Sigma_{200}-i\Sigma_{100}\right).
\end{align}
This model has a particle-hole symmetry represented by $U_{\mathcal P}=\Sigma_{001}$, which is diagonalized by $ M_{\mathcal P}=\frac{1}{\sqrt{2}}\left(\Sigma_{000}+i\Sigma_{002}\right)$. In real space, this transformation again changes basis from a Dirac to a Majorana basis. The spectrum of a finite 3D model with open boundary conditions, shown in \Cref{fig:cubeAdjMaj}(c), displays a gapped bulk along with a pair of zero-energy states.
These correspond to a pair of perfectly localized Majorana corner modes.

\section{Conclusion}
In this work, we presented three models hosting exact Majorana corner states at zero energy, and show how to identify the model parameters that result in perfectly localized modes based on the intersections of nontrivial solutions of common null-vectors of all hopping matrices. Inspired by the dimerization structure revealed in a particular 2D model, we demonstrated that 2D models with pairs of topological Majorana corner modes can be constructed from sets of coupled Kitaev chains. We proposed how to extend the unit cell to stabilize localized corner modes in different geometric configurations and to embed them in higher-dimensional bulk models.
We further identified a new topological invariant -- the nested Pfaffian -- which is uniquely linked to the geometric configuration of the pair of corner modes.
Finally, by proposing a pumping protocol which adiabatically connects different configurations of corner modes, we showed how to adiabatically move localized Majorana states in a finite-sized lattice.
In a previous work~\cite{PhysRevResearch.2.032068}, the pumping protocol was proposed for braiding of Majorana bound states. While the possible outcomes of statistical phases obtained through such braiding protocols are obstructed by the geometry, an extension to 3D with two pairs of Majorana corner modes could overcome such limitations.

\acknowledgments
PPP acknowledges support by the Laboratory for Physical Sciences through the Condensed Matter Theory Center. AH acknowledges financial support from the Luxembourg National Research Fund under grant C20/MS/14764976/TOPREL. PPP thanks Katharina Laubscher for helpful discussions.

\begin{appendix}
\onecolumngrid

\section{Diagonalising the Particle-Hole Symmetry}\label{sec:appA}
In this section, we discuss the construction of the Majorana basis of the DIII model discussed in \cref{eq:2D_adjacent_corners}. The unitary part of the particle hole symmetry is given by $U_\mathcal P = \Sigma_{31}$, so that the Hamiltonian satisfies ${H}_{\textbf{k}}= -\Sigma_{31}{H}_{-\textbf{k}}^*\Sigma_{31}$. Along with particle-hole symmetry, there also exists the chiral symmetry $U_\mathcal C=\Sigma_{21}$ satisfying ${H}_{\textbf{k}}= -\Sigma_{21}{H}_{\textbf{k}}\Sigma_{21}$. $U_\mathcal P$ and $U_\mathcal C$ can be simultaneously diagonalized using the matrix $M_\mathcal P=\exp[i\frac{\pi}{4}\left(\Sigma_{30}-\Sigma_{03}\right)]\exp[i\frac{\pi}{4}\left(\Sigma_{02}-\Sigma_{10}\right)]$, such that
\begin{align}
   M_{\mathcal P} U_\mathcal C M_\mathcal P^\dagger &= \text{Diag }(1,-1,-1,1)\\
   M_{\mathcal P} U_\mathcal P M_\mathcal P^T &= \text{Diag }(1,1,1,1).
\end{align}
The Matrix $M_\mathcal P$ is given by
\begin{align}
M_\mathcal P = \frac12\begin{pmatrix}1&0&0&0\\0&i&0&0\\0&0&-i&0\\0&0&0&1\end{pmatrix}\begin{pmatrix}1&-i&1&-i\\-i&1&-i&1\\-1&i&1&-i\\i&-1&-i&1\end{pmatrix}.
\end{align}
The Hamiltonian is defined over the basis $\bm c_{r} = (c_{\uparrow,{\bm r}}, c_{\downarrow,{\bm r}}, c_{\uparrow,{\bm r}}^\dagger,- c_{\downarrow,{\bm r}}^\dagger)^T$. Under change of basis by $M_\mathcal P$, the new basis in real space are Majorana fermions, given by
\begin{align}
\begin{pmatrix}
    m_{{\bm r},1}\\m_{{\bm r},2}\\m_{{\bm r},3}\\m_{{\bm r},4}
\end{pmatrix} &= M_\mathcal P \begin{pmatrix}
    c_{\uparrow,{\bm r}}\\ c_{\downarrow,{\bm r}}\\c_{\uparrow,{\bm r}}^\dagger\\- c_{\downarrow,{\bm r}}^\dagger
\end{pmatrix} \notag \\
&= \frac12 \begin{pmatrix}
    c_{\uparrow,{\bm r}}+ c_{\uparrow,{\bm r}}^\dagger -i(c_{\downarrow,{\bm r}} - c_{\downarrow,{\bm r}}^\dagger)\\
   c_{\uparrow,{\bm r}}+ c_{\uparrow,{\bm r}}^\dagger +i(c_{\downarrow,{\bm r}} - c_{\downarrow,{\bm r}}^\dagger)\\
   i( c_{\uparrow,{\bm r}} -c_{\uparrow,{\bm r}}^\dagger)+ c_{\downarrow,{\bm r}} +c_{\downarrow,{\bm r}}^\dagger\\
    i (c_{\uparrow,{\bm r}} - c_{\uparrow,{\bm r}}^\dagger)- (c_{\downarrow,{\bm r}}  +c_{\downarrow,{\bm r}}^\dagger)
\end{pmatrix}.
\end{align}
The Majorana basis defined here satisfies $m_{{\bm r},i}^\dagger=m_{{\bm r},i}$, and they anticommute with each other, i.e. $\{m_{\bm r,i},m_{\bm r,j}\}=\frac12\delta_{ij}$.

\section{Perfectly Localized limit of DIII SOTI }\label{sec:appB}
The model that hosts perfectly localized corner states is inspired by the Hamiltonian in Ref.~\cite{PhysRevResearch.2.032068}. In this work, the Authors study a Hamiltonian of the DIII topological class with counter-propagating edge helical Majorana edge states. By introducing an in-plane magnetic field, the edge states hybridized into a pair of Majorana corner states.

\begin{figure}[t]
    \centering
    \includegraphics{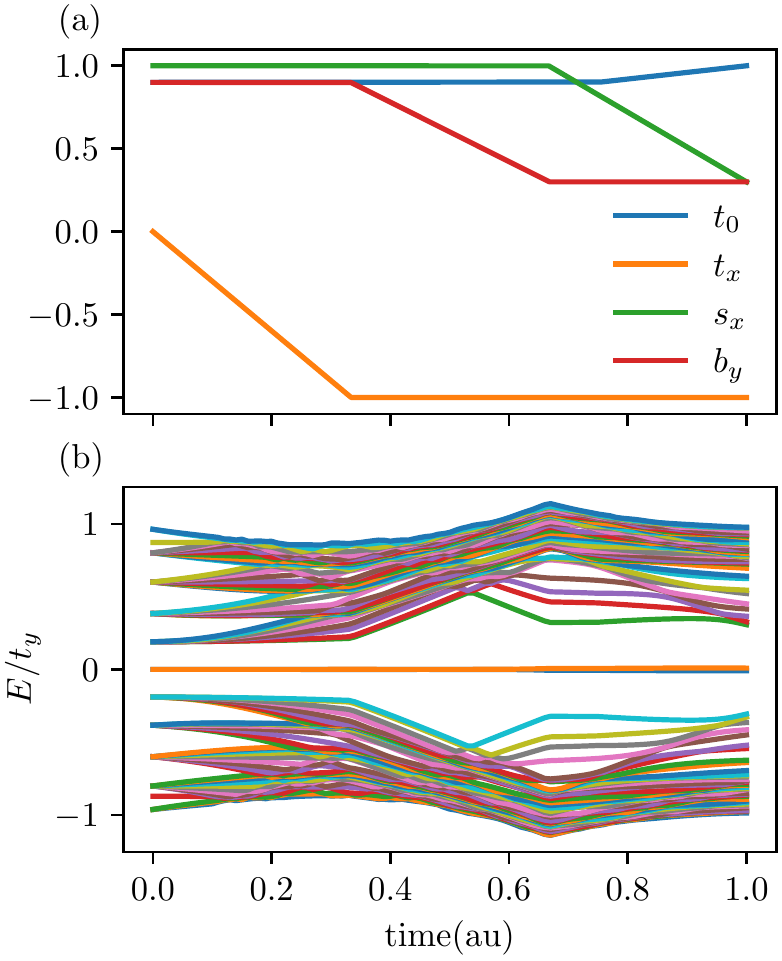}
    \caption{(a) The change in the parameters as the function of the deformation parameters (b) The spectrum of the corresponding open system along the deformation process. Neither the bulk nor the edge gap closes, and the zero energy states remain at zero energy. Results shown for a $7\times 7$ system size.   }
    \label{fig:DeformationToPahomiFig_deformparam}
\end{figure}

To establish an adiabatic connection, we need to define a deformation process that interpolates between \cref{eq:2D_adjacent_corners} and the model studied in Ref.~\cite{PhysRevResearch.2.032068}, which has the Hamiltonian
\begin{align}
{H^G}_{\bm k} \label{eq:HPah}
&=
   t_0 (1-\cos k_x-\cos k_y)\Sigma_{03}\\
&-t_0\sin k_y \Sigma_{02} +t_0\sin k_x\Sigma_{31}\notag\\
&+
    (s_x\cos k_x +s_y\cos k_y)\Sigma_{11}+ b_x\Sigma_{10}+b_y\Sigma_{23},\notag
\end{align}
while the generic interpolating Hamiltonian is given by
\begin{align}
{H}_{\bm k} \label{eq:Hgen}
&=
    (t_0+t_x\cos k_x+t_y\cos k_y)\Sigma_{03}\\
&+
    d_{y}\sin k_y \Sigma_{02} +d_{x}\sin k_x\Sigma_{31}\notag\\
&+
    (s_x\cos k_x +s_y\cos k_y)\Sigma_{11}+ b_x\Sigma_{10}+b_y\Sigma_{23},\notag.
\end{align}
We start with the following parameters for the interpolating Hamiltonian,
\begin{align}
    t_x&=b_x=0.0, & b_y&=t_0=0.9, \notag \\
    d_y&=t_y=-1.0,& d_x&=s_x=1.0.
\end{align}
The corresponding Hamiltonian hosts perfectly localized Majorana modes at $r=(0,0)$ and $r=(0,L)$, as described in the main text. Next, we deform the parameters $t_0,t_x,s_x,b_y$, keeping the remaining parameters constant, via the deformation process shown in \cref{fig:DeformationToPahomiFig_deformparam}(a), to reach the final state defined by
\begin{align}
    t_x&=-1.0,& b_x&=0.0, & b_y& =0.3,\notag \\
    t_0&=1.0, &d_y& =-1.0,& d_x& =1.0,& s_x&=0.3,
\end{align}
which corresponds to the Hamiltonian \ref{eq:HPah}. Through this process, we show the spectrum of the open boundary finite size system in \cref{fig:DeformationToPahomiFig_deformparam}(b). The bulk and edge gap remains nonzero during the deformation, which concludes the demonstration that the two models are adiabatically connected.

\section{The Edge Spectrum of the DIII Model and the Pfaffian}\label{sec:appC}

\begin{figure}[t]
    \centering
    \includegraphics[scale=0.5]{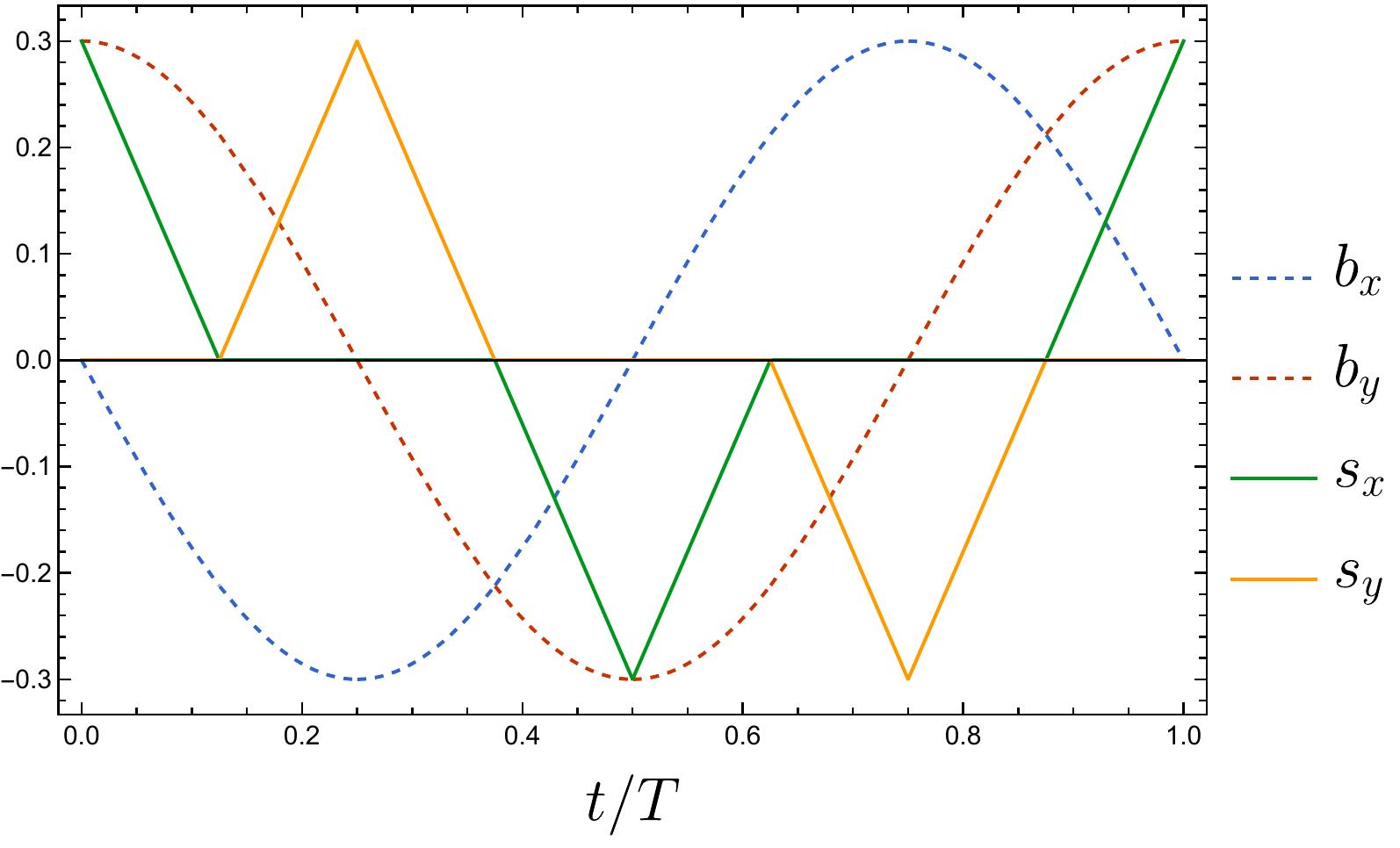}
    \caption{The parameters of the pump process described in Ref.~\cite{PhysRevResearch.2.032068}, in units of $t_0$, over a full cycle. }
    \label{fig:pumpPahomi}
\end{figure}

In this section, we consider the ribbon geometry of the two-dimensional model hosting Majorana corner modes on adjacent corners, as described by \cref{eq:ribbondef}, for the Hamiltonian in \cref{eq:2D_adjacent_corners}. Our goal is to show that the Pfaffian invariant introduced in the main text accurately describes the topological phases of the edge, by considering the pump process of Ref.~\cite{PhysRevResearch.2.032068}. The pump process uses the Hamiltonian \ref{eq:HPah}, with the parameters of the cycle shown in \cref{fig:pumpPahomi}.

\begin{figure}[t]
    \centering
    \includegraphics[scale=0.5]{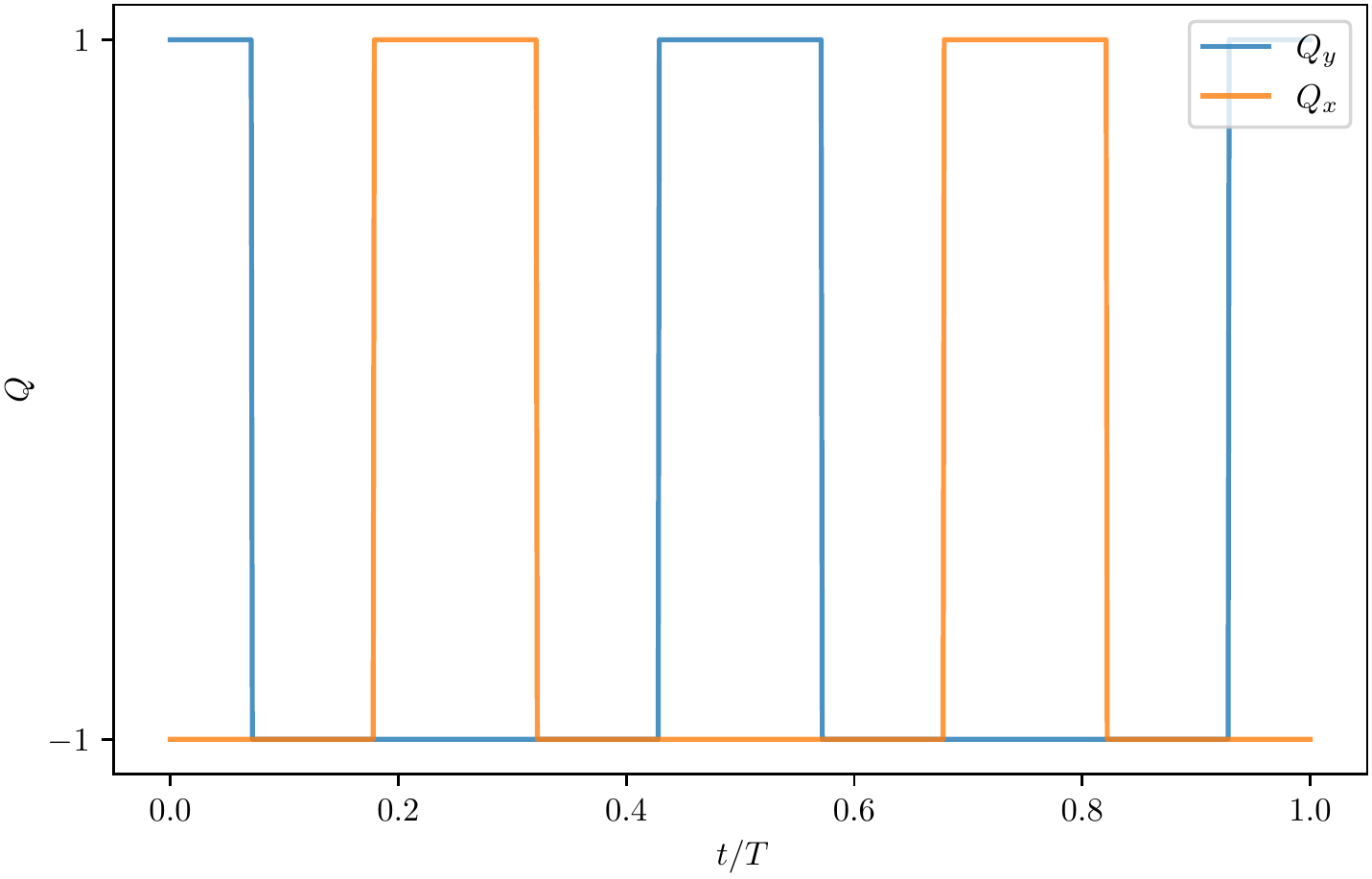}
    \caption{The pair of Pfaffians along the pump process of Ref.~\cite{PhysRevResearch.2.032068}, accurately capturing the topology of the edge model. Results shown for a $7\times 7$ system size.}
    \label{fig:pfaffian}
\end{figure}

In \cref{fig:pfaffian}(a), we show the Pfaffian of Hamiltonian through the pump cycle as a function of the pump time $t$. As we can see, at $t=0$ we start with the $x$ edge topological and $y$ edge trivial. As one of the Majorana is pumped through the $y$ edge, it becomes topological at around $t_1/T\sim 0.075$. Next, the $x$ edge gap closes and the $x$ edge becomes trivial at around $t_2/T\sim 0.185$, and the process repeats. In \cref{fig:pfaffian}(b), we show the corresponding ribbon geometry spectrum at $t=0,t_1,t_2$. At $t=0$, the edge spectrum along $x$ and $y$ are both gapped. At $t=t_1$, the $y$ edge gap closes, which changes the topology of the $y$ edge from trivial to non-trivial. At $t=t_2$, the $x$ edge gap closes, changing the $x$ edge topology to a trivial state.

\begin{figure}[t]
    \centering
    \includegraphics[scale=0.5]{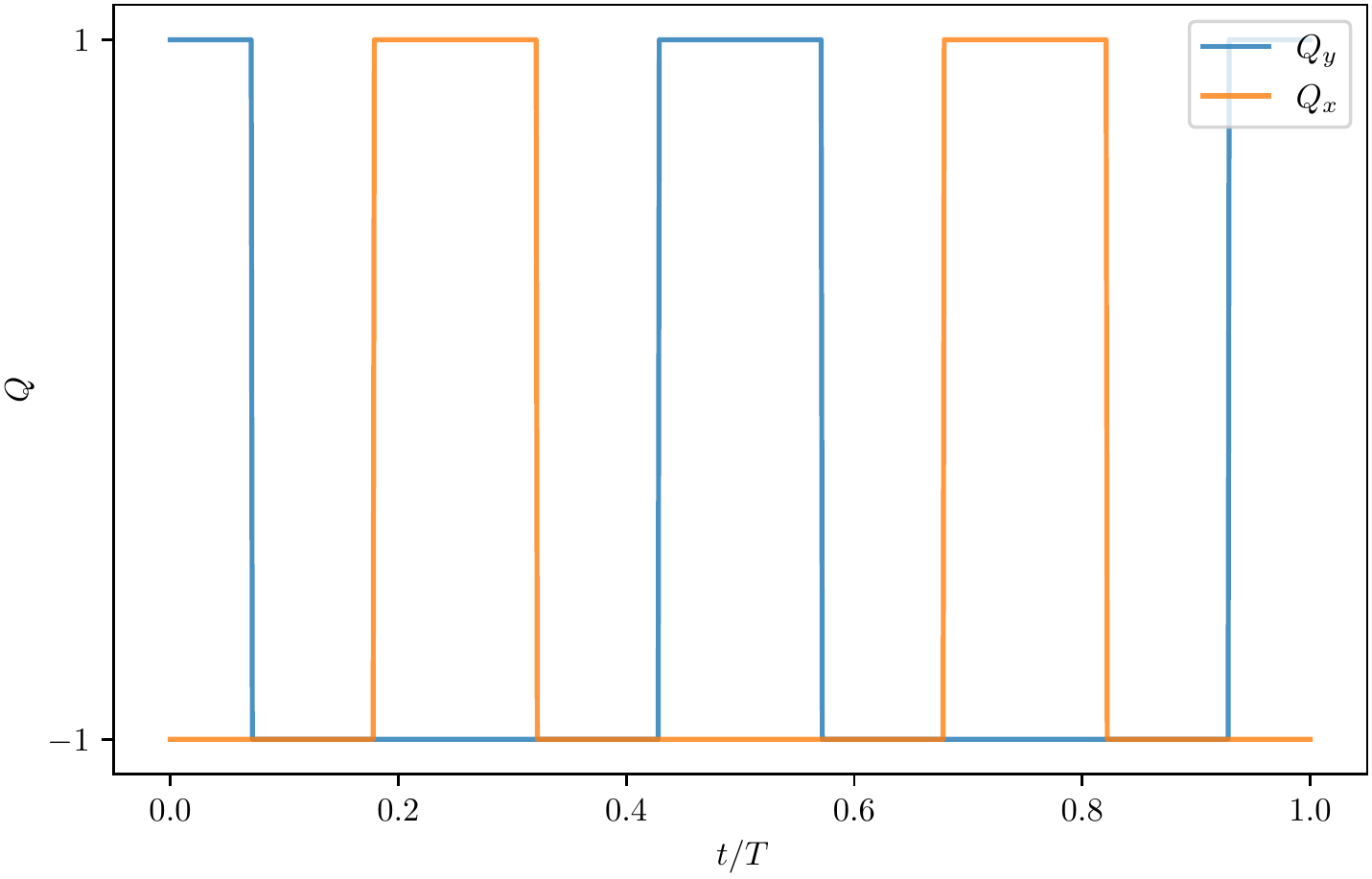}
    \caption{The pair of Pfaffians along the pump process of Ref.~\cite{PhysRevResearch.2.032068}, accurately capturing the topology of the edge model. Results shown for a $7\times 7$ system size.}
    \label{fig:pfaffian}
\end{figure}

\begin{figure}[t]
    \centering
    \includegraphics[scale=0.75]{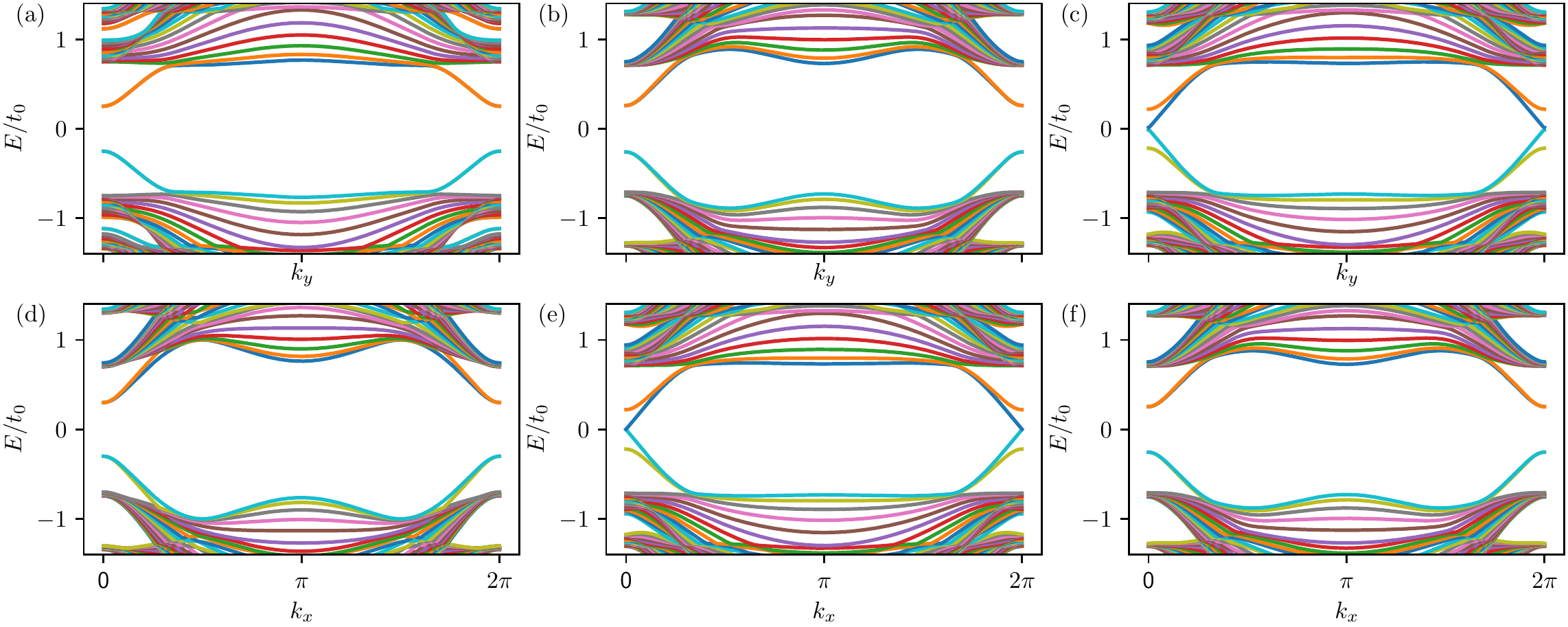}
    \caption{The edge spectra from the ribbon geometry during the pump process of Ref.~\cite{PhysRevResearch.2.032068}, (a)-(c) with a finite width along $x$ and periodic along $y$ and (d)-(f) vice versa. For (a),(d) it is at $t=0$ when the $x$ edge is topological and $y$ edge is trivial. (b),(e) At $t/T\sim 0.075$, the $y$ edge gap closes and becomes topological and (c),(f) at $t_2/T\sim 0.185$, the $x$ edge gap closes and becomes trivial.}
    \label{fig:ribbon}
\end{figure}

\section{Bulk spectrum in the 2D pumping procedure}\label{sec:appD}
\begin{figure}[ht]
    \centering
    \includegraphics[scale=0.55]{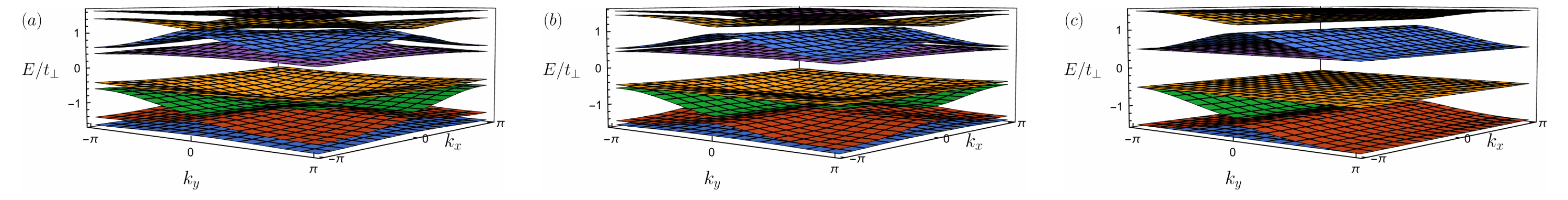}
    \caption{The bulk spectra of \cref{eq:bilayer_T} during the 2D pumping procedure described in \cref{eq:pump2D} at $t=0,0.5,1$.}
    \label{fig:bulk2Dpump}
\end{figure}

In \cref{fig:bulk2Dpump}, we show the bulk spectrum of the Hamiltonian of \cref{eq:bilayer_T} during the 2D pumping procedure described in \cref{eq:pump2D}, at $t=0,0.5,1$ of the pumping process. We see that the bulk spectrum is clearly gapped at all the shown times, including at $t=0.5$ which is when the edge gap closes and the topology changes.

\section{Ribbon and face spectra}\label{sec:appE}

In this section, we show the ribbon and face spectra for the 2D adjacent Majorana model, and for the corresponding 3D model.

\begin{figure}[ht]
    \centering
    \includegraphics[scale=0.75]{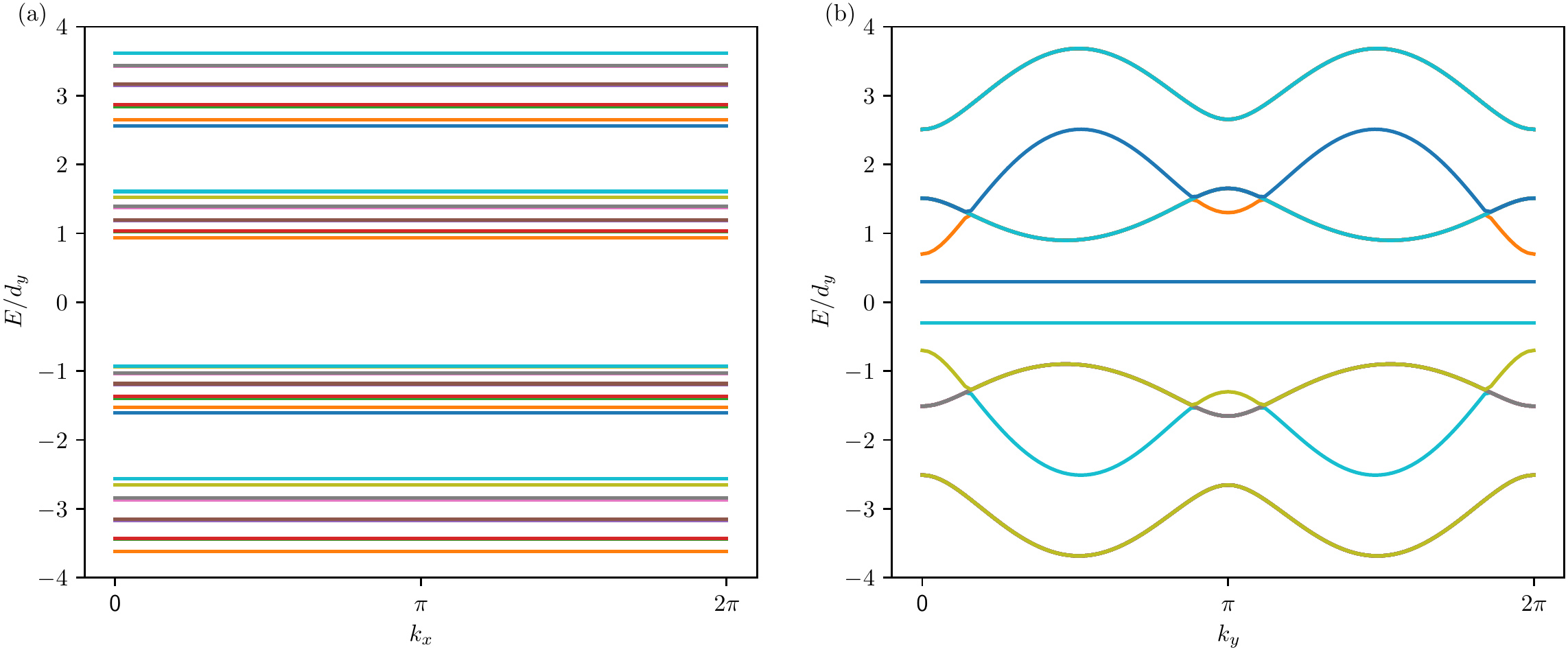}
    \caption{The edge spectra from the ribbon geometry for the original $DIII$ model (a) with a finite width along $y$ and periodic along $x$ and (b) vice versa.}
    \label{fig:ribbon2Dadj}
\end{figure}

\Cref{fig:ribbon2Dadj}(a) shows the spectrum for the ribbon model of the 2D model which is finite along $y$ and periodic along $x$. This shows us the spectra of the $x-$edge and consists completely of flat bands. This is because the model is dimerized along the $x$ edge, consisting of units that extend along the $y-$edge but decoupled along the $x$ direction. \Cref{fig:ribbon2Dadj}(b) shows the corresponding spectrum along the $y-$edge, which is dispersive as expected. The two lowest bands are also flat, however the energy at which they lie can be tuned by changing $t_y$.

\begin{figure}[ht]
    \centering
    \includegraphics[scale=0.55]{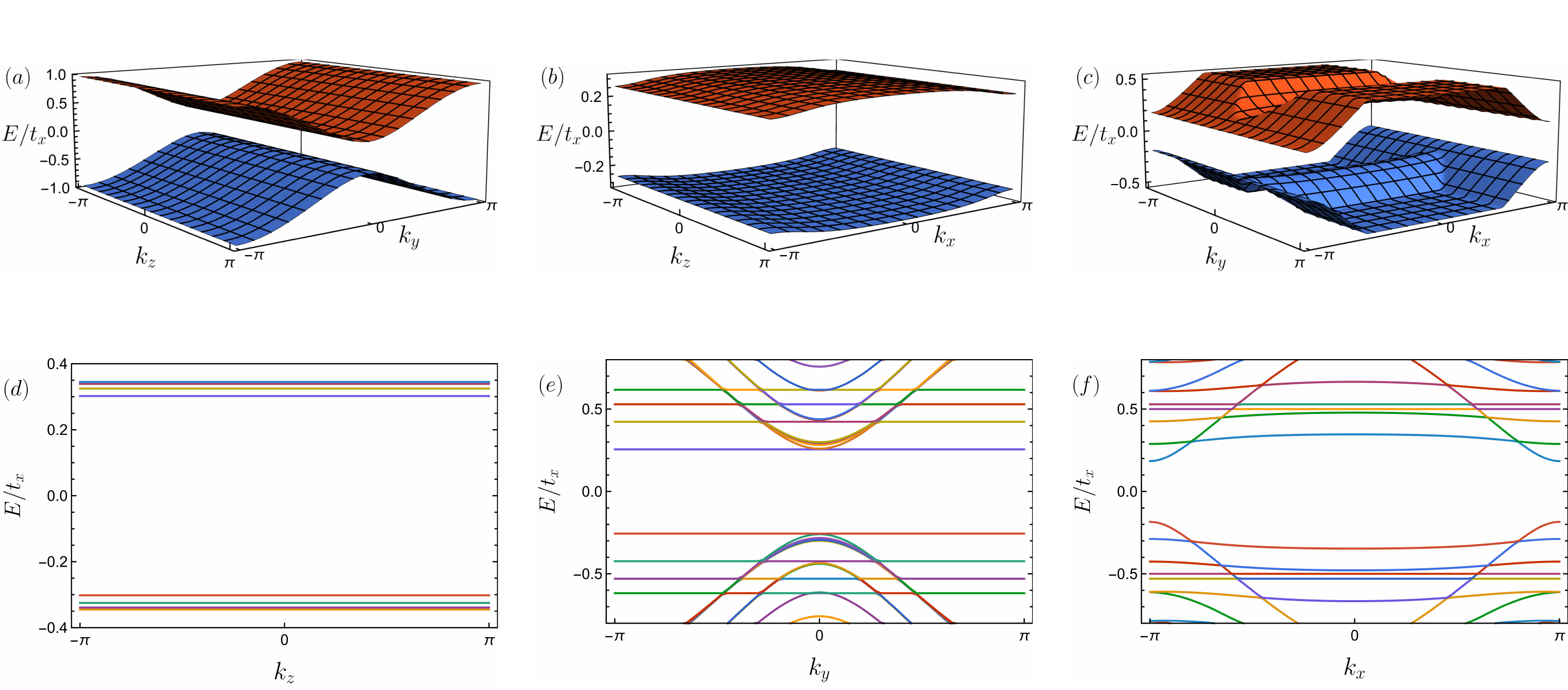}
    \caption{ The boundary spectra of the 3D model (a)-(c)The face spectra (d)-(e) the edge spectra. Results shown for a $7\times7\times7$ system size with parameters $\alpha_1=-0.9t_x, \mu_1=\mu_2=-0.25t_x,t_{y}=t'_{y}=0.45t_x,t_z=-1.2t_x$. }
    \label{fig:3Dfaceribbon}
\end{figure}

Next, we look at the face and ribbon spectrum of the 3D model consisting of adjacent Majorana corner modes. The face spectrum is constructed by considering a system that is finite in the direction perpendicular to the face, and periodic along the directions that define the face. \Cref{fig:3Dfaceribbon}(a-c) shows the face spectrum for the face perpendicular to $x,y,z$ direction respectively. We see that for the $yz$ face, the bands are not dispersive along the $z$ direction. Moreover, the bands of the $zx$ face are approximately flat (but not exactly). \Cref{fig:3Dfaceribbon}(d-f) shows the ribbon spectrum along the $z,y,x$ direction respectively. The ribbon band spectrum should be related to the face bands by projecting the appropriate face spectra along the appropriate directions, modulo finite size effects. We see that all the face and edge ribbon spectra are gapped.

\twocolumngrid
\end{appendix}

\clearpage

\bibliography{refs.bib}

\begin{thebibliography}{38}%
\makeatletter
\providecommand \@ifxundefined [1]{%
 \@ifx{#1\undefined}
}%
\providecommand \@ifnum [1]{%
 \ifnum #1\expandafter \@firstoftwo
 \else \expandafter \@secondoftwo
 \fi
}%
\providecommand \@ifx [1]{%
 \ifx #1\expandafter \@firstoftwo
 \else \expandafter \@secondoftwo
 \fi
}%
\providecommand \natexlab [1]{#1}%
\providecommand \enquote  [1]{``#1''}%
\providecommand \bibnamefont  [1]{#1}%
\providecommand \bibfnamefont [1]{#1}%
\providecommand \citenamefont [1]{#1}%
\providecommand \href@noop [0]{\@secondoftwo}%
\providecommand \href [0]{\begingroup \@sanitize@url \@href}%
\providecommand \@href[1]{\@@startlink{#1}\@@href}%
\providecommand \@@href[1]{\endgroup#1\@@endlink}%
\providecommand \@sanitize@url [0]{\catcode `\\12\catcode `\$12\catcode
  `\&12\catcode `\#12\catcode `\^12\catcode `\_12\catcode `\%12\relax}%
\providecommand \@@startlink[1]{}%
\providecommand \@@endlink[0]{}%
\providecommand \url  [0]{\begingroup\@sanitize@url \@url }%
\providecommand \@url [1]{\endgroup\@href {#1}{\urlprefix }}%
\providecommand \urlprefix  [0]{URL }%
\providecommand \Eprint [0]{\href }%
\providecommand \doibase [0]{https://doi.org/}%
\providecommand \selectlanguage [0]{\@gobble}%
\providecommand \bibinfo  [0]{\@secondoftwo}%
\providecommand \bibfield  [0]{\@secondoftwo}%
\providecommand \translation [1]{[#1]}%
\providecommand \BibitemOpen [0]{}%
\providecommand \bibitemStop [0]{}%
\providecommand \bibitemNoStop [0]{.\EOS\space}%
\providecommand \EOS [0]{\spacefactor3000\relax}%
\providecommand \BibitemShut  [1]{\csname bibitem#1\endcsname}%
\let\auto@bib@innerbib\@empty
%</preamble>
\bibitem [{\citenamefont {Phong}\ \emph {et~al.}(2017)\citenamefont {Phong},
  \citenamefont {Walet},\ and\ \citenamefont {Guinea}}]{phong2017majorana}%
  \BibitemOpen
  \bibfield  {author} {\bibinfo {author} {\bibfnamefont {V.~T.}\ \bibnamefont
  {Phong}}, \bibinfo {author} {\bibfnamefont {N.~R.}\ \bibnamefont {Walet}},\
  and\ \bibinfo {author} {\bibfnamefont {F.}~\bibnamefont {Guinea}},\
  }\bibfield  {title} {\bibinfo {title} {Majorana zero modes in a
  two-dimensional p-wave superconductor},\ }\href@noop {} {\bibfield  {journal}
  {\bibinfo  {journal} {Physical Review B}\ }\textbf {\bibinfo {volume} {96}},\
  \bibinfo {pages} {060505} (\bibinfo {year} {2017})}\BibitemShut {NoStop}%
\bibitem [{\citenamefont {Pahomi}\ \emph {et~al.}(2020)\citenamefont {Pahomi},
  \citenamefont {Sigrist},\ and\ \citenamefont
  {Soluyanov}}]{PhysRevResearch.2.032068}%
  \BibitemOpen
  \bibfield  {author} {\bibinfo {author} {\bibfnamefont {T.~E.}\ \bibnamefont
  {Pahomi}}, \bibinfo {author} {\bibfnamefont {M.}~\bibnamefont {Sigrist}},\
  and\ \bibinfo {author} {\bibfnamefont {A.~A.}\ \bibnamefont {Soluyanov}},\
  }\bibfield  {title} {\bibinfo {title} {Braiding majorana corner modes in a
  second-order topological superconductor},\ }\href@noop {} {\bibfield
  {journal} {\bibinfo  {journal} {Phyical Review Research}\ }\textbf {\bibinfo
  {volume} {2}},\ \bibinfo {pages} {032068} (\bibinfo {year}
  {2020})}\BibitemShut {NoStop}%
\bibitem [{\citenamefont {Ryu}\ \emph {et~al.}(2010)\citenamefont {Ryu},
  \citenamefont {Schnyder}, \citenamefont {Furusaki},\ and\ \citenamefont
  {Ludwig}}]{ryu2010topological}%
  \BibitemOpen
  \bibfield  {author} {\bibinfo {author} {\bibfnamefont {S.}~\bibnamefont
  {Ryu}}, \bibinfo {author} {\bibfnamefont {A.~P.}\ \bibnamefont {Schnyder}},
  \bibinfo {author} {\bibfnamefont {A.}~\bibnamefont {Furusaki}},\ and\
  \bibinfo {author} {\bibfnamefont {A.~W.}\ \bibnamefont {Ludwig}},\ }\bibfield
   {title} {\bibinfo {title} {Topological insulators and superconductors:
  tenfold way and dimensional hierarchy},\ }\href@noop {} {\bibfield  {journal}
  {\bibinfo  {journal} {New Journal of Physics}\ }\textbf {\bibinfo {volume}
  {12}},\ \bibinfo {pages} {065010} (\bibinfo {year} {2010})}\BibitemShut
  {NoStop}%
\bibitem [{\citenamefont {Laubscher}\ \emph
  {et~al.}(2020{\natexlab{a}})\citenamefont {Laubscher}, \citenamefont
  {Chughtai}, \citenamefont {Loss},\ and\ \citenamefont
  {Klinovaja}}]{laubscher2020kramers}%
  \BibitemOpen
  \bibfield  {author} {\bibinfo {author} {\bibfnamefont {K.}~\bibnamefont
  {Laubscher}}, \bibinfo {author} {\bibfnamefont {D.}~\bibnamefont {Chughtai}},
  \bibinfo {author} {\bibfnamefont {D.}~\bibnamefont {Loss}},\ and\ \bibinfo
  {author} {\bibfnamefont {J.}~\bibnamefont {Klinovaja}},\ }\bibfield  {title}
  {\bibinfo {title} {Kramers pairs of majorana corner states in a topological
  insulator bilayer},\ }\href@noop {} {\bibfield  {journal} {\bibinfo
  {journal} {Physical Review B}\ }\textbf {\bibinfo {volume} {102}},\ \bibinfo
  {pages} {195401} (\bibinfo {year} {2020}{\natexlab{a}})}\BibitemShut
  {NoStop}%
\bibitem [{\citenamefont {Laubscher}\ \emph
  {et~al.}(2020{\natexlab{b}})\citenamefont {Laubscher}, \citenamefont {Loss},\
  and\ \citenamefont {Klinovaja}}]{laubscher2020majorana}%
  \BibitemOpen
  \bibfield  {author} {\bibinfo {author} {\bibfnamefont {K.}~\bibnamefont
  {Laubscher}}, \bibinfo {author} {\bibfnamefont {D.}~\bibnamefont {Loss}},\
  and\ \bibinfo {author} {\bibfnamefont {J.}~\bibnamefont {Klinovaja}},\
  }\bibfield  {title} {\bibinfo {title} {Majorana and parafermion corner states
  from two coupled sheets of bilayer graphene},\ }\href@noop {} {\bibfield
  {journal} {\bibinfo  {journal} {Physical Review Research}\ }\textbf {\bibinfo
  {volume} {2}},\ \bibinfo {pages} {013330} (\bibinfo {year}
  {2020}{\natexlab{b}})}\BibitemShut {NoStop}%
\bibitem [{\citenamefont {Laubscher}\ and\ \citenamefont
  {Klinovaja}(2021)}]{laubscher2021majorana}%
  \BibitemOpen
  \bibfield  {author} {\bibinfo {author} {\bibfnamefont {K.}~\bibnamefont
  {Laubscher}}\ and\ \bibinfo {author} {\bibfnamefont {J.}~\bibnamefont
  {Klinovaja}},\ }\bibfield  {title} {\bibinfo {title} {Majorana bound states
  in semiconducting nanostructures},\ }\href@noop {} {\bibfield  {journal}
  {\bibinfo  {journal} {Journal of Applied Physics}\ }\textbf {\bibinfo
  {volume} {130}},\ \bibinfo {pages} {081101} (\bibinfo {year}
  {2021})}\BibitemShut {NoStop}%
\bibitem [{\citenamefont {Hasan}\ and\ \citenamefont
  {Kane}(2010)}]{hasan2010colloquium}%
  \BibitemOpen
  \bibfield  {author} {\bibinfo {author} {\bibfnamefont {M.~Z.}\ \bibnamefont
  {Hasan}}\ and\ \bibinfo {author} {\bibfnamefont {C.~L.}\ \bibnamefont
  {Kane}},\ }\bibfield  {title} {\bibinfo {title} {Colloquium: topological
  insulators},\ }\href@noop {} {\bibfield  {journal} {\bibinfo  {journal}
  {Reviews of modern physics}\ }\textbf {\bibinfo {volume} {82}},\ \bibinfo
  {pages} {3045} (\bibinfo {year} {2010})}\BibitemShut {NoStop}%
\bibitem [{\citenamefont {Chiu}\ \emph {et~al.}(2016)\citenamefont {Chiu},
  \citenamefont {Teo}, \citenamefont {Schnyder},\ and\ \citenamefont
  {Ryu}}]{chiu2016classification}%
  \BibitemOpen
  \bibfield  {author} {\bibinfo {author} {\bibfnamefont {C.-K.}\ \bibnamefont
  {Chiu}}, \bibinfo {author} {\bibfnamefont {J.~C.}\ \bibnamefont {Teo}},
  \bibinfo {author} {\bibfnamefont {A.~P.}\ \bibnamefont {Schnyder}},\ and\
  \bibinfo {author} {\bibfnamefont {S.}~\bibnamefont {Ryu}},\ }\bibfield
  {title} {\bibinfo {title} {Classification of topological quantum matter with
  symmetries},\ }\href@noop {} {\bibfield  {journal} {\bibinfo  {journal}
  {Reviews of Modern Physics}\ }\textbf {\bibinfo {volume} {88}},\ \bibinfo
  {pages} {035005} (\bibinfo {year} {2016})}\BibitemShut {NoStop}%
\bibitem [{\citenamefont {Moore}(2010)}]{moore2010birth}%
  \BibitemOpen
  \bibfield  {author} {\bibinfo {author} {\bibfnamefont {J.~E.}\ \bibnamefont
  {Moore}},\ }\bibfield  {title} {\bibinfo {title} {The birth of topological
  insulators},\ }\href@noop {} {\bibfield  {journal} {\bibinfo  {journal}
  {Nature}\ }\textbf {\bibinfo {volume} {464}},\ \bibinfo {pages} {194}
  (\bibinfo {year} {2010})}\BibitemShut {NoStop}%
\bibitem [{\citenamefont {Benalcazar}\ \emph
  {et~al.}(2017{\natexlab{a}})\citenamefont {Benalcazar}, \citenamefont
  {Bernevig},\ and\ \citenamefont {Hughes}}]{benalcazar2017electric}%
  \BibitemOpen
  \bibfield  {author} {\bibinfo {author} {\bibfnamefont {W.~A.}\ \bibnamefont
  {Benalcazar}}, \bibinfo {author} {\bibfnamefont {B.~A.}\ \bibnamefont
  {Bernevig}},\ and\ \bibinfo {author} {\bibfnamefont {T.~L.}\ \bibnamefont
  {Hughes}},\ }\bibfield  {title} {\bibinfo {title} {Electric multipole
  moments, topological multipole moment pumping, and chiral hinge states in
  crystalline insulators},\ }\href@noop {} {\bibfield  {journal} {\bibinfo
  {journal} {Physical Review B}\ }\textbf {\bibinfo {volume} {96}},\ \bibinfo
  {pages} {245115} (\bibinfo {year} {2017}{\natexlab{a}})}\BibitemShut
  {NoStop}%
\bibitem [{\citenamefont {Benalcazar}\ \emph
  {et~al.}(2017{\natexlab{b}})\citenamefont {Benalcazar}, \citenamefont
  {Bernevig},\ and\ \citenamefont {Hughes}}]{benalcazar2017quantized}%
  \BibitemOpen
  \bibfield  {author} {\bibinfo {author} {\bibfnamefont {W.~A.}\ \bibnamefont
  {Benalcazar}}, \bibinfo {author} {\bibfnamefont {B.~A.}\ \bibnamefont
  {Bernevig}},\ and\ \bibinfo {author} {\bibfnamefont {T.~L.}\ \bibnamefont
  {Hughes}},\ }\bibfield  {title} {\bibinfo {title} {Quantized electric
  multipole insulators},\ }\href@noop {} {\bibfield  {journal} {\bibinfo
  {journal} {Science}\ }\textbf {\bibinfo {volume} {357}},\ \bibinfo {pages}
  {61} (\bibinfo {year} {2017}{\natexlab{b}})}\BibitemShut {NoStop}%
\bibitem [{\citenamefont {Imhof}\ \emph {et~al.}(2018)\citenamefont {Imhof},
  \citenamefont {Berger}, \citenamefont {Bayer}, \citenamefont {Brehm},
  \citenamefont {Molenkamp}, \citenamefont {Kiessling}, \citenamefont
  {Schindler}, \citenamefont {Lee}, \citenamefont {Greiter}, \citenamefont
  {Neupert} \emph {et~al.}}]{imhof2018topolectrical}%
  \BibitemOpen
  \bibfield  {author} {\bibinfo {author} {\bibfnamefont {S.}~\bibnamefont
  {Imhof}}, \bibinfo {author} {\bibfnamefont {C.}~\bibnamefont {Berger}},
  \bibinfo {author} {\bibfnamefont {F.}~\bibnamefont {Bayer}}, \bibinfo
  {author} {\bibfnamefont {J.}~\bibnamefont {Brehm}}, \bibinfo {author}
  {\bibfnamefont {L.~W.}\ \bibnamefont {Molenkamp}}, \bibinfo {author}
  {\bibfnamefont {T.}~\bibnamefont {Kiessling}}, \bibinfo {author}
  {\bibfnamefont {F.}~\bibnamefont {Schindler}}, \bibinfo {author}
  {\bibfnamefont {C.~H.}\ \bibnamefont {Lee}}, \bibinfo {author} {\bibfnamefont
  {M.}~\bibnamefont {Greiter}}, \bibinfo {author} {\bibfnamefont
  {T.}~\bibnamefont {Neupert}}, \emph {et~al.},\ }\bibfield  {title} {\bibinfo
  {title} {Topolectrical-circuit realization of topological corner modes},\
  }\href@noop {} {\bibfield  {journal} {\bibinfo  {journal} {Nature Physics}\
  }\textbf {\bibinfo {volume} {14}},\ \bibinfo {pages} {925} (\bibinfo {year}
  {2018})}\BibitemShut {NoStop}%
\bibitem [{\citenamefont {Geier}\ \emph {et~al.}(2018)\citenamefont {Geier},
  \citenamefont {Trifunovic}, \citenamefont {Hoskam},\ and\ \citenamefont
  {Brouwer}}]{geier2018second}%
  \BibitemOpen
  \bibfield  {author} {\bibinfo {author} {\bibfnamefont {M.}~\bibnamefont
  {Geier}}, \bibinfo {author} {\bibfnamefont {L.}~\bibnamefont {Trifunovic}},
  \bibinfo {author} {\bibfnamefont {M.}~\bibnamefont {Hoskam}},\ and\ \bibinfo
  {author} {\bibfnamefont {P.~W.}\ \bibnamefont {Brouwer}},\ }\bibfield
  {title} {\bibinfo {title} {Second-order topological insulators and
  superconductors with an order-two crystalline symmetry},\ }\href@noop {}
  {\bibfield  {journal} {\bibinfo  {journal} {Physical Review B}\ }\textbf
  {\bibinfo {volume} {97}},\ \bibinfo {pages} {205135} (\bibinfo {year}
  {2018})}\BibitemShut {NoStop}%
\bibitem [{\citenamefont {Schindler}\ \emph {et~al.}(2018)\citenamefont
  {Schindler}, \citenamefont {Cook}, \citenamefont {Vergniory}, \citenamefont
  {Wang}, \citenamefont {Parkin}, \citenamefont {Bernevig},\ and\ \citenamefont
  {Neupert}}]{schindler2018higher}%
  \BibitemOpen
  \bibfield  {author} {\bibinfo {author} {\bibfnamefont {F.}~\bibnamefont
  {Schindler}}, \bibinfo {author} {\bibfnamefont {A.~M.}\ \bibnamefont {Cook}},
  \bibinfo {author} {\bibfnamefont {M.~G.}\ \bibnamefont {Vergniory}}, \bibinfo
  {author} {\bibfnamefont {Z.}~\bibnamefont {Wang}}, \bibinfo {author}
  {\bibfnamefont {S.~S.}\ \bibnamefont {Parkin}}, \bibinfo {author}
  {\bibfnamefont {B.~A.}\ \bibnamefont {Bernevig}},\ and\ \bibinfo {author}
  {\bibfnamefont {T.}~\bibnamefont {Neupert}},\ }\bibfield  {title} {\bibinfo
  {title} {Higher-order topological insulators},\ }\href@noop {} {\bibfield
  {journal} {\bibinfo  {journal} {Science advances}\ }\textbf {\bibinfo
  {volume} {4}},\ \bibinfo {pages} {eaat0346} (\bibinfo {year}
  {2018})}\BibitemShut {NoStop}%
\bibitem [{\citenamefont {Benalcazar}\ \emph {et~al.}(2014)\citenamefont
  {Benalcazar}, \citenamefont {Teo},\ and\ \citenamefont
  {Hughes}}]{benalcazar2014classification}%
  \BibitemOpen
  \bibfield  {author} {\bibinfo {author} {\bibfnamefont {W.~A.}\ \bibnamefont
  {Benalcazar}}, \bibinfo {author} {\bibfnamefont {J.~C.}\ \bibnamefont
  {Teo}},\ and\ \bibinfo {author} {\bibfnamefont {T.~L.}\ \bibnamefont
  {Hughes}},\ }\bibfield  {title} {\bibinfo {title} {Classification of
  two-dimensional topological crystalline superconductors and majorana bound
  states at disclinations},\ }\href@noop {} {\bibfield  {journal} {\bibinfo
  {journal} {Physical Review B}\ }\textbf {\bibinfo {volume} {89}},\ \bibinfo
  {pages} {224503} (\bibinfo {year} {2014})}\BibitemShut {NoStop}%
\bibitem [{\citenamefont {Song}\ \emph {et~al.}(2017)\citenamefont {Song},
  \citenamefont {Fang},\ and\ \citenamefont {Fang}}]{song2017d}%
  \BibitemOpen
  \bibfield  {author} {\bibinfo {author} {\bibfnamefont {Z.}~\bibnamefont
  {Song}}, \bibinfo {author} {\bibfnamefont {Z.}~\bibnamefont {Fang}},\ and\
  \bibinfo {author} {\bibfnamefont {C.}~\bibnamefont {Fang}},\ }\bibfield
  {title} {\bibinfo {title} {(d- 2)-dimensional edge states of rotation
  symmetry protected topological states},\ }\href@noop {} {\bibfield  {journal}
  {\bibinfo  {journal} {Physical review letters}\ }\textbf {\bibinfo {volume}
  {119}},\ \bibinfo {pages} {246402} (\bibinfo {year} {2017})}\BibitemShut
  {NoStop}%
\bibitem [{\citenamefont {Schnyder}\ \emph {et~al.}(2008)\citenamefont
  {Schnyder}, \citenamefont {Ryu}, \citenamefont {Furusaki},\ and\
  \citenamefont {Ludwig}}]{schnyder2008classification}%
  \BibitemOpen
  \bibfield  {author} {\bibinfo {author} {\bibfnamefont {A.~P.}\ \bibnamefont
  {Schnyder}}, \bibinfo {author} {\bibfnamefont {S.}~\bibnamefont {Ryu}},
  \bibinfo {author} {\bibfnamefont {A.}~\bibnamefont {Furusaki}},\ and\
  \bibinfo {author} {\bibfnamefont {A.~W.}\ \bibnamefont {Ludwig}},\ }\bibfield
   {title} {\bibinfo {title} {Classification of topological insulators and
  superconductors in three spatial dimensions},\ }\href@noop {} {\bibfield
  {journal} {\bibinfo  {journal} {Physical Review B}\ }\textbf {\bibinfo
  {volume} {78}},\ \bibinfo {pages} {195125} (\bibinfo {year}
  {2008})}\BibitemShut {NoStop}%
\bibitem [{\citenamefont {Fu}(2011)}]{fu2011topological}%
  \BibitemOpen
  \bibfield  {author} {\bibinfo {author} {\bibfnamefont {L.}~\bibnamefont
  {Fu}},\ }\bibfield  {title} {\bibinfo {title} {Topological crystalline
  insulators},\ }\href@noop {} {\bibfield  {journal} {\bibinfo  {journal}
  {Physical Review Letters}\ }\textbf {\bibinfo {volume} {106}},\ \bibinfo
  {pages} {106802} (\bibinfo {year} {2011})}\BibitemShut {NoStop}%
\bibitem [{\citenamefont {Zhang}\ \emph {et~al.}(2019)\citenamefont {Zhang},
  \citenamefont {Wang}, \citenamefont {Lin}, \citenamefont {Tian},
  \citenamefont {Xie}, \citenamefont {Lu}, \citenamefont {Chen},\ and\
  \citenamefont {Jiang}}]{zhang2019second}%
  \BibitemOpen
  \bibfield  {author} {\bibinfo {author} {\bibfnamefont {X.}~\bibnamefont
  {Zhang}}, \bibinfo {author} {\bibfnamefont {H.-X.}\ \bibnamefont {Wang}},
  \bibinfo {author} {\bibfnamefont {Z.-K.}\ \bibnamefont {Lin}}, \bibinfo
  {author} {\bibfnamefont {Y.}~\bibnamefont {Tian}}, \bibinfo {author}
  {\bibfnamefont {B.}~\bibnamefont {Xie}}, \bibinfo {author} {\bibfnamefont
  {M.-H.}\ \bibnamefont {Lu}}, \bibinfo {author} {\bibfnamefont {Y.-F.}\
  \bibnamefont {Chen}},\ and\ \bibinfo {author} {\bibfnamefont {J.-H.}\
  \bibnamefont {Jiang}},\ }\bibfield  {title} {\bibinfo {title} {Second-order
  topology and multidimensional topological transitions in sonic crystals},\
  }\href@noop {} {\bibfield  {journal} {\bibinfo  {journal} {Nature Physics}\
  }\textbf {\bibinfo {volume} {15}},\ \bibinfo {pages} {582} (\bibinfo {year}
  {2019})}\BibitemShut {NoStop}%
\bibitem [{\citenamefont {Nguyen}\ \emph {et~al.}(2022)\citenamefont {Nguyen},
  \citenamefont {Brzezicki},\ and\ \citenamefont {Hyart}}]{nguyen2022corner}%
  \BibitemOpen
  \bibfield  {author} {\bibinfo {author} {\bibfnamefont {N.~M.}\ \bibnamefont
  {Nguyen}}, \bibinfo {author} {\bibfnamefont {W.}~\bibnamefont {Brzezicki}},\
  and\ \bibinfo {author} {\bibfnamefont {T.}~\bibnamefont {Hyart}},\ }\bibfield
   {title} {\bibinfo {title} {Corner states, hinge states, and majorana modes
  in snte nanowires},\ }\href@noop {} {\bibfield  {journal} {\bibinfo
  {journal} {Physical Review B}\ }\textbf {\bibinfo {volume} {105}},\ \bibinfo
  {pages} {075310} (\bibinfo {year} {2022})}\BibitemShut {NoStop}%
\bibitem [{\citenamefont {Zhang}\ and\ \citenamefont
  {Sarma}(2021)}]{zhang2021intrinsic}%
  \BibitemOpen
  \bibfield  {author} {\bibinfo {author} {\bibfnamefont {R.-X.}\ \bibnamefont
  {Zhang}}\ and\ \bibinfo {author} {\bibfnamefont {S.~D.}\ \bibnamefont
  {Sarma}},\ }\bibfield  {title} {\bibinfo {title} {Intrinsic
  time-reversal-invariant topological superconductivity in thin films of
  iron-based superconductors},\ }\href@noop {} {\bibfield  {journal} {\bibinfo
  {journal} {Physical Review Letters}\ }\textbf {\bibinfo {volume} {126}},\
  \bibinfo {pages} {137001} (\bibinfo {year} {2021})}\BibitemShut {NoStop}%
\bibitem [{\citenamefont {Pan}\ \emph {et~al.}(2021)\citenamefont {Pan},
  \citenamefont {Luo}, \citenamefont {Gao},\ and\ \citenamefont
  {Liu}}]{pan2021braiding}%
  \BibitemOpen
  \bibfield  {author} {\bibinfo {author} {\bibfnamefont {X.-H.}\ \bibnamefont
  {Pan}}, \bibinfo {author} {\bibfnamefont {X.-J.}\ \bibnamefont {Luo}},
  \bibinfo {author} {\bibfnamefont {J.-H.}\ \bibnamefont {Gao}},\ and\ \bibinfo
  {author} {\bibfnamefont {X.}~\bibnamefont {Liu}},\ }\bibfield  {title}
  {\bibinfo {title} {Braiding higher-order majorana corner states through their
  spin degree of freedom},\ }\href@noop {} {\bibfield  {journal} {\bibinfo
  {journal} {arXiv preprint arXiv:2111.12359}\ } (\bibinfo {year}
  {2021})}\BibitemShut {NoStop}%
\bibitem [{\citenamefont {Ezawa}(2018{\natexlab{a}})}]{ezawa2018higher}%
  \BibitemOpen
  \bibfield  {author} {\bibinfo {author} {\bibfnamefont {M.}~\bibnamefont
  {Ezawa}},\ }\bibfield  {title} {\bibinfo {title} {Higher-order topological
  insulators and semimetals on the breathing kagome and pyrochlore lattices},\
  }\href@noop {} {\bibfield  {journal} {\bibinfo  {journal} {Physical review
  letters}\ }\textbf {\bibinfo {volume} {120}},\ \bibinfo {pages} {026801}
  (\bibinfo {year} {2018}{\natexlab{a}})}\BibitemShut {NoStop}%
\bibitem [{\citenamefont {Laubscher}\ \emph {et~al.}(2019)\citenamefont
  {Laubscher}, \citenamefont {Loss},\ and\ \citenamefont
  {Klinovaja}}]{laubscher2019fractional}%
  \BibitemOpen
  \bibfield  {author} {\bibinfo {author} {\bibfnamefont {K.}~\bibnamefont
  {Laubscher}}, \bibinfo {author} {\bibfnamefont {D.}~\bibnamefont {Loss}},\
  and\ \bibinfo {author} {\bibfnamefont {J.}~\bibnamefont {Klinovaja}},\
  }\bibfield  {title} {\bibinfo {title} {Fractional topological
  superconductivity and parafermion corner states},\ }\href@noop {} {\bibfield
  {journal} {\bibinfo  {journal} {Physical Review Research}\ }\textbf {\bibinfo
  {volume} {1}},\ \bibinfo {pages} {032017} (\bibinfo {year}
  {2019})}\BibitemShut {NoStop}%
\bibitem [{\citenamefont {Laubscher}\ \emph {et~al.}(2023)\citenamefont
  {Laubscher}, \citenamefont {Keizer},\ and\ \citenamefont
  {Klinovaja}}]{laubscher2023fractional}%
  \BibitemOpen
  \bibfield  {author} {\bibinfo {author} {\bibfnamefont {K.}~\bibnamefont
  {Laubscher}}, \bibinfo {author} {\bibfnamefont {P.}~\bibnamefont {Keizer}},\
  and\ \bibinfo {author} {\bibfnamefont {J.}~\bibnamefont {Klinovaja}},\
  }\bibfield  {title} {\bibinfo {title} {Fractional second-order topological
  insulator from a three-dimensional coupled-wires construction},\ }\href@noop
  {} {\bibfield  {journal} {\bibinfo  {journal} {Physical Review B}\ }\textbf
  {\bibinfo {volume} {107}},\ \bibinfo {pages} {045409} (\bibinfo {year}
  {2023})}\BibitemShut {NoStop}%
\bibitem [{\citenamefont {Kitaev}(2001)}]{kitaev2001unpaired}%
  \BibitemOpen
  \bibfield  {author} {\bibinfo {author} {\bibfnamefont {A.~Y.}\ \bibnamefont
  {Kitaev}},\ }\bibfield  {title} {\bibinfo {title} {Unpaired majorana fermions
  in quantum wires},\ }\href@noop {} {\bibfield  {journal} {\bibinfo  {journal}
  {Physics-uspekhi}\ }\textbf {\bibinfo {volume} {44}},\ \bibinfo {pages} {131}
  (\bibinfo {year} {2001})}\BibitemShut {NoStop}%
\bibitem [{\citenamefont {Budich}\ and\ \citenamefont
  {Ardonne}(2013)}]{budich2013equivalent}%
  \BibitemOpen
  \bibfield  {author} {\bibinfo {author} {\bibfnamefont {J.~C.}\ \bibnamefont
  {Budich}}\ and\ \bibinfo {author} {\bibfnamefont {E.}~\bibnamefont
  {Ardonne}},\ }\bibfield  {title} {\bibinfo {title} {Equivalent topological
  invariants for one-dimensional majorana wires in symmetry class d},\
  }\href@noop {} {\bibfield  {journal} {\bibinfo  {journal} {Physical Review
  B}\ }\textbf {\bibinfo {volume} {88}},\ \bibinfo {pages} {075419} (\bibinfo
  {year} {2013})}\BibitemShut {NoStop}%
\bibitem [{\citenamefont {Peng}\ \emph {et~al.}(2017)\citenamefont {Peng},
  \citenamefont {Bao},\ and\ \citenamefont {von Oppen}}]{peng2017boundary}%
  \BibitemOpen
  \bibfield  {author} {\bibinfo {author} {\bibfnamefont {Y.}~\bibnamefont
  {Peng}}, \bibinfo {author} {\bibfnamefont {Y.}~\bibnamefont {Bao}},\ and\
  \bibinfo {author} {\bibfnamefont {F.}~\bibnamefont {von Oppen}},\ }\bibfield
  {title} {\bibinfo {title} {Boundary green functions of topological insulators
  and superconductors},\ }\href@noop {} {\bibfield  {journal} {\bibinfo
  {journal} {Physical Review B}\ }\textbf {\bibinfo {volume} {95}},\ \bibinfo
  {pages} {235143} (\bibinfo {year} {2017})}\BibitemShut {NoStop}%
\bibitem [{\citenamefont {Zhang}\ \emph {et~al.}(2020)\citenamefont {Zhang},
  \citenamefont {Sau},\ and\ \citenamefont {Sarma}}]{zhang2020kitaev}%
  \BibitemOpen
  \bibfield  {author} {\bibinfo {author} {\bibfnamefont {R.-X.}\ \bibnamefont
  {Zhang}}, \bibinfo {author} {\bibfnamefont {J.~D.}\ \bibnamefont {Sau}},\
  and\ \bibinfo {author} {\bibfnamefont {S.~D.}\ \bibnamefont {Sarma}},\
  }\bibfield  {title} {\bibinfo {title} {Kitaev building-block construction for
  higher-order topological superconductors},\ }\href@noop {} {\bibfield
  {journal} {\bibinfo  {journal} {arXiv preprint arXiv:2003.02559}\ } (\bibinfo
  {year} {2020})}\BibitemShut {NoStop}%
\bibitem [{\citenamefont {Ezawa}(2018{\natexlab{b}})}]{ezawa2018minimal}%
  \BibitemOpen
  \bibfield  {author} {\bibinfo {author} {\bibfnamefont {M.}~\bibnamefont
  {Ezawa}},\ }\bibfield  {title} {\bibinfo {title} {Minimal models for
  wannier-type higher-order topological insulators and phosphorene},\
  }\href@noop {} {\bibfield  {journal} {\bibinfo  {journal} {Physical Review
  B}\ }\textbf {\bibinfo {volume} {98}},\ \bibinfo {pages} {045125} (\bibinfo
  {year} {2018}{\natexlab{b}})}\BibitemShut {NoStop}%
\bibitem [{\citenamefont {Khalaf}(2018)}]{khalaf2018higher}%
  \BibitemOpen
  \bibfield  {author} {\bibinfo {author} {\bibfnamefont {E.}~\bibnamefont
  {Khalaf}},\ }\bibfield  {title} {\bibinfo {title} {Higher-order topological
  insulators and superconductors protected by inversion symmetry},\ }\href@noop
  {} {\bibfield  {journal} {\bibinfo  {journal} {Physical Review B}\ }\textbf
  {\bibinfo {volume} {97}},\ \bibinfo {pages} {205136} (\bibinfo {year}
  {2018})}\BibitemShut {NoStop}%
\bibitem [{\citenamefont {Bradlyn}\ \emph {et~al.}(2017)\citenamefont
  {Bradlyn}, \citenamefont {Elcoro}, \citenamefont {Cano}, \citenamefont
  {Vergniory}, \citenamefont {Wang}, \citenamefont {Felser}, \citenamefont
  {Aroyo},\ and\ \citenamefont {Bernevig}}]{bradlyn2017topological}%
  \BibitemOpen
  \bibfield  {author} {\bibinfo {author} {\bibfnamefont {B.}~\bibnamefont
  {Bradlyn}}, \bibinfo {author} {\bibfnamefont {L.}~\bibnamefont {Elcoro}},
  \bibinfo {author} {\bibfnamefont {J.}~\bibnamefont {Cano}}, \bibinfo {author}
  {\bibfnamefont {M.~G.}\ \bibnamefont {Vergniory}}, \bibinfo {author}
  {\bibfnamefont {Z.}~\bibnamefont {Wang}}, \bibinfo {author} {\bibfnamefont
  {C.}~\bibnamefont {Felser}}, \bibinfo {author} {\bibfnamefont {M.~I.}\
  \bibnamefont {Aroyo}},\ and\ \bibinfo {author} {\bibfnamefont {B.~A.}\
  \bibnamefont {Bernevig}},\ }\bibfield  {title} {\bibinfo {title} {Topological
  quantum chemistry},\ }\href@noop {} {\bibfield  {journal} {\bibinfo
  {journal} {Nature}\ }\textbf {\bibinfo {volume} {547}},\ \bibinfo {pages}
  {298} (\bibinfo {year} {2017})}\BibitemShut {NoStop}%
\bibitem [{\citenamefont {Simon}\ \emph {et~al.}(2022)\citenamefont {Simon},
  \citenamefont {Geier},\ and\ \citenamefont {Brouwer}}]{simon2022higher}%
  \BibitemOpen
  \bibfield  {author} {\bibinfo {author} {\bibfnamefont {S.}~\bibnamefont
  {Simon}}, \bibinfo {author} {\bibfnamefont {M.}~\bibnamefont {Geier}},\ and\
  \bibinfo {author} {\bibfnamefont {P.~W.}\ \bibnamefont {Brouwer}},\
  }\bibfield  {title} {\bibinfo {title} {Higher-order topological semimetals
  and nodal superconductors with an order-two crystalline symmetry},\
  }\href@noop {} {\bibfield  {journal} {\bibinfo  {journal} {Physical Review
  B}\ }\textbf {\bibinfo {volume} {106}},\ \bibinfo {pages} {035105} (\bibinfo
  {year} {2022})}\BibitemShut {NoStop}%
\bibitem [{\citenamefont {Trifunovic}\ and\ \citenamefont
  {Brouwer}(2021)}]{trifunovic2021higher}%
  \BibitemOpen
  \bibfield  {author} {\bibinfo {author} {\bibfnamefont {L.}~\bibnamefont
  {Trifunovic}}\ and\ \bibinfo {author} {\bibfnamefont {P.~W.}\ \bibnamefont
  {Brouwer}},\ }\bibfield  {title} {\bibinfo {title} {Higher-order topological
  band structures},\ }\href@noop {} {\bibfield  {journal} {\bibinfo  {journal}
  {Physica Status Solidi (b)}\ }\textbf {\bibinfo {volume} {258}},\ \bibinfo
  {pages} {2000090} (\bibinfo {year} {2021})}\BibitemShut {NoStop}%
\bibitem [{\citenamefont {Hu}\ and\ \citenamefont
  {Zhang}(2023)}]{hu2023topological}%
  \BibitemOpen
  \bibfield  {author} {\bibinfo {author} {\bibfnamefont {L.-H.}\ \bibnamefont
  {Hu}}\ and\ \bibinfo {author} {\bibfnamefont {R.-X.}\ \bibnamefont {Zhang}},\
  }\bibfield  {title} {\bibinfo {title} {Topological superconducting vortex
  from trivial electronic bands},\ }\href@noop {} {\bibfield  {journal}
  {\bibinfo  {journal} {Nature Communications}\ }\textbf {\bibinfo {volume}
  {14}},\ \bibinfo {pages} {640} (\bibinfo {year} {2023})}\BibitemShut
  {NoStop}%
\bibitem [{\citenamefont {Lenggenhager}\ \emph {et~al.}(2022)\citenamefont
  {Lenggenhager}, \citenamefont {Liu}, \citenamefont {Neupert},\ and\
  \citenamefont {Bzdu{\v{s}}ek}}]{lenggenhager2022universal}%
  \BibitemOpen
  \bibfield  {author} {\bibinfo {author} {\bibfnamefont {P.~M.}\ \bibnamefont
  {Lenggenhager}}, \bibinfo {author} {\bibfnamefont {X.}~\bibnamefont {Liu}},
  \bibinfo {author} {\bibfnamefont {T.}~\bibnamefont {Neupert}},\ and\ \bibinfo
  {author} {\bibfnamefont {T.}~\bibnamefont {Bzdu{\v{s}}ek}},\ }\bibfield
  {title} {\bibinfo {title} {Universal higher-order bulk-boundary
  correspondence of triple nodal points},\ }\href@noop {} {\bibfield  {journal}
  {\bibinfo  {journal} {Physical Review B}\ }\textbf {\bibinfo {volume}
  {106}},\ \bibinfo {pages} {085129} (\bibinfo {year} {2022})}\BibitemShut
  {NoStop}%
\bibitem [{\citenamefont {Fidkowski}\ \emph {et~al.}(2011)\citenamefont
  {Fidkowski}, \citenamefont {Jackson},\ and\ \citenamefont
  {Klich}}]{fidkowski2011model}%
  \BibitemOpen
  \bibfield  {author} {\bibinfo {author} {\bibfnamefont {L.}~\bibnamefont
  {Fidkowski}}, \bibinfo {author} {\bibfnamefont {T.}~\bibnamefont {Jackson}},\
  and\ \bibinfo {author} {\bibfnamefont {I.}~\bibnamefont {Klich}},\ }\bibfield
   {title} {\bibinfo {title} {Model characterization of gapless edge modes of
  topological insulators using intermediate brillouin-zone functions},\
  }\href@noop {} {\bibfield  {journal} {\bibinfo  {journal} {Physical review
  letters}\ }\textbf {\bibinfo {volume} {107}},\ \bibinfo {pages} {036601}
  (\bibinfo {year} {2011})}\BibitemShut {NoStop}%
\bibitem [{\citenamefont {Matsugatani}\ and\ \citenamefont
  {Watanabe}(2018)}]{matsugatani2018connecting}%
  \BibitemOpen
  \bibfield  {author} {\bibinfo {author} {\bibfnamefont {A.}~\bibnamefont
  {Matsugatani}}\ and\ \bibinfo {author} {\bibfnamefont {H.}~\bibnamefont
  {Watanabe}},\ }\bibfield  {title} {\bibinfo {title} {Connecting higher-order
  topological insulators to lower-dimensional topological insulators},\
  }\href@noop {} {\bibfield  {journal} {\bibinfo  {journal} {Physical Review
  B}\ }\textbf {\bibinfo {volume} {98}},\ \bibinfo {pages} {205129} (\bibinfo
  {year} {2018})}\BibitemShut {NoStop}%
\end{thebibliography}%

\end{document}